\documentclass[12pt]{article}
\usepackage{amssymb}
  \makeatletter
\renewcommand\thesection{\@Roman\c@section}
\renewcommand\thesubsection{\thesection.\@arabic\c@subsection}
\makeatother

\newcommand{\sect}[1]{\setcounter{equation}{0}\section{#1}}
\renewcommand{\theequation}{\thesection.\arabic{equation}}

\begin{document}
\begin{center}
{\Large \bf Drinfeld Twists and Algebraic Bethe Ansatz of the
Supersymmetric $t$-$J$ Model } \vskip.2in {\large Wen-Li
Yang$^{1,2}$, Yao-Zhong Zhang$^{2}$ and Shao-You Zhao$^{2,3}$
 } \vskip.2in
 {\em $^{1}$ Institute of Modern Physics, Northwest University,
Xi'an 710069, China}\\
{\em $^{2}$ Department of Mathematics, University of
Queensland, Brisbane 4072, Australia }\\

 {\em $^{3}$  Department of Physics, Beijing Institute of
Technology, Beijing 100081, China }\vskip.1in {\em E-mail:}
wenli@maths.uq.edu.au, yzz@maths.uq.edu.au, syz@maths.uq.edu.au

\end{center}
\begin{abstract}
We construct the Drinfeld twists (factorizing $F$-matrices) for
the supersymmetric $t$-$J$ model. Working in the basis provided by
the $F$-matrix (i.e. the so-called $F$-basis), we obtain
completely symmetric representations of the monodromy matrix and
the pseudo-particle creation operators of the model. These enable
us to resolve the hierarchy of the nested Bethe vectors for the
$gl(2|1)$ invariant $t$-$J$ model.
\end{abstract}

\sect{Introduction} ~~~ The algebraic Bethe ansatz or the quantum
inverse scattering method (QISM) provides a powerful tool of
solving eigenvalue problems such as diagonalizing integrable
two-dimensional quantum spin chains. In this framework, the
pseudo-particle creation and annihilation operators are
constructed by the off-diagonal entries of the monodromy matrix.
The Bethe vectors (eigenvectors) are obtained by acting the
creation operators on the pseudo-vacuum state. However, the
apparently simple action of creation operators is intricate on the
level of the local operators by non-local effects arising from
polarization clouds or compensating exchange terms. This makes the
exact and explicit computation of correlation functions difficult
(if not impossible).

Recently, Maillet and Sanchez de Santos \cite{Maillet96} showed
how monodromy matrices of the inhomogeneous XXX and XXZ spin
chains can be simplified by using the factorizing Drinfeld twists.
This leads to the natural $F$-basis for the analysis of these
models. In this basis, the pseudo-particle creation and
annihilation operators take completely symmetric forms and contain
no compensating exchange terms on the level of the local operators
(i.e. polarization free). As a result, the Bethe vectors of the
models are simplified dramatically and can be written down
explicitly.

The results of \cite{Maillet96} were generalized to certain other
systems. In \cite{Terras99}, the Drinfeld twists associated with
any finite-dimensional irreducible representations of the Yangian
$Y[gl(2)]$ were investigated. In \cite{Kitanine98}, the form
factors for local spin operators of the spin-1/2 XXZ model were
computed and in \cite{Izergin98}, the spontaneous magnetization of
the XXZ chain on the finite lattice was represented. In
\cite{Albert00}, Albert et al constructed the $F$-matrix of the
$gl(m)$ rational Heisenberg model and obtained a polarization free
representation of the creation operators. Using these results,
they resolved the hierarchy of the nested Bethe ansatz for the
$gl(m)$ model. In \cite{Albert0005}\cite{Albert0007}, the Drinfeld
twists of the elliptic XYZ model and Belavin model were
constructed.

The $t$-$J$ model was proposed in an attempt to understand
high-$T_c$ superconductivity \cite{Suth75,Schultz83,Wieg88,And90}.
It is a strongly correlated electron system with nearest-neighbor
hopping ($t$) and anti-ferromagnetic exchange ($J$) of electrons.
When $J=2t$, the $t$-$J$ model becomes $gl(2|1)$ invariant. Using
the nested algebraic Bethe ansatz method, Essler and Korepin
obtained the eigenvalues of the supersymmetric $t$-$J$ model
\cite{Ess92}. The algebraic structure and physical properties of
the model were investigated in \cite{Foer931,Bergere01,Ambjorn01}.

In this paper, we construct the factorizing $F$-matrix of the
supersymmetric $t$-$J$ model. Working in the $F$-basis, we obtain
the symmetric representations of the monodromy matrix and the
creation operators. Using these results, we resolve the hierarchy
of the nested Bethe vectors of the $gl(2|1)$ invariant $t$-$J$
model.

The present paper is organized as follows. In section 2, we
introduce some basic notation of the supersymmetric $t$-$J$ model.
In section 3, we construct the $F$-matrix and its inverse. In
section 4, we give the representation of the monodromy matrix and
the creation operators in the $F$-basis. The nested Bethe vectors
of the model are resolved in section 5. We conclude the paper by
offering some discussions in section 6.

\sect{Basic definitions and notation}
 ~~~
Let $V$ be the 3-dimensional $gl(2|1)$-module and $R\in
End(V\otimes V)$ the $R$-matrix associated with this module. $V$
is $Z_2$-graded, and in the following we choose the BBF grading
for $V$, i.e. $[1]=[2]=0,[3]=1$. The $R$-matrix depends on the
difference of two spectral parameters $u_1$ and $u_2$ associated
with the two copies of $V$, and is, in the BBF grading, given by
\begin{eqnarray}
 R_{12}(u_1,u_2)=R_{12}(u_1-u_2)%\nonumber\\
% &=& \nonumber\\
          &=&\left(\begin{array}{ccccccccc}
 1&0&0 & 0&0&0 & 0&0&0\\
 0&a_{12}&0 & b_{12}&0&0 & 0&0&0 \\
 0&0&a_{12} & 0&0&0  & b_{12}&0&0\\
 0&b_{12}&0 & a_{12}&0&0  & 0&0&0\\
 0&0&0      & 0&1&0  & 0&0&0\\
 0&0&0      & 0&0&a_{12}  & 0&b_{12}&0\\
 0&0&b_{12} & 0&0&0       & a_{12}&0&0\\
 0&0&0      & 0&0&b_{12}  & 0&a_{12}&0\\
 0&0&0      & 0&0&0       & 0&0&c_{12}
 \end{array} \right), \label{de:R} \nonumber\\
\end{eqnarray}
where
\begin{eqnarray}
&& a_{12}=a(u_1,u_2)\equiv {u_1-u_2\over u_1-u_2+\eta},\quad \quad
b_{12}=b(u_1,u_2)\equiv{\eta\over u_1-u_2+\eta},\quad\quad
\nonumber\\
&& c_{12}=c(u_1,u_2)\equiv{u_1-u_2-\eta\over u_1-u_2+\eta}
\end{eqnarray}
with $\eta\in C$ being the crossing parameter. One can easily
check that the $R$-matrix satisfies the unitary relation
\begin{equation}
R_{21}R_{12}=1.
\end{equation}
Here and throughout $R_{12}\equiv R_{12}(u_1,u_2)$. The $R$-matrix
satisfies the graded Yang-Baxter equation (GYBE)
\begin{equation}
R_{12}R_{13}R_{23}=R_{23}R_{13}R_{12}.
\end{equation}
In terms of the matrix elements defined by
\begin{equation}
R(u)(v^{i'}\otimes v^{j'})=\sum_{i,j}R(u)^{i'j'}_{ij}(v^{i}\otimes
v^{j}),
\end{equation}
the GYBE reads
\begin{eqnarray}
&&
\sum_{i',j',k'}R(u_1-u_2)^{i'j'}_{ij}R(u_1-u_3)^{i''k'}_{i'k}R(u_2-u_3)^{j''k''}_{j'k'}
    (-1)^{[j']([i']+[i''])}\nonumber\\
&=&\sum_{i',j',k'}R(u_2-u_3)^{j'k'}_{jk}R(u_1-u_3)^{i'k''}_{ik'}R(u_1-u_2)^{i''j''}_{i'j'}
    (-1)^{[j']([i]+[i'])}.
\end{eqnarray}

The quantum monodromy matrix $T(u)$ of the supersymmetric $t$-$J$
chain of length $N$ is defined as
\begin{eqnarray}
T(u)=R_{0N}(u,z_N)R_{0N-1}(u,z_{N-1})_{\ldots} R_{01}(u,z_1),
\label{de:T}
\end{eqnarray}
where the index 0 refers to the auxiliary space  and $\{z_i\}$ are
arbitrary inhomogeneous parameters depending on site $i$. $T(u)$
can be represented in the auxiliary space as the $3\times 3$
matrix whose elements are operators acting on the quantum space
$V^{\otimes N}$:
\begin{eqnarray}
 {T}(u)=\left(\begin{array}{ccc}
{ A}_{11}(u)& {A}_{12}(u)&{ B}_1(u)\\
{ A}_{21}(u)& { A}_{22}(u)&{ B}_2(u)\\
{ C}_{1}(u)& { C}_{2}(u)&{ D}(u)
 \end{array}\right). \label{de:T-marix}
 \end{eqnarray}
 By using the GYBE, one may prove that the monodromy matrix
satisfies the GYBE
\begin{eqnarray}
R_{12}(u-v)T_1(u)T_2(v)=T_2(v)T_1(u)R_{12}(u-v).\label{eq:GYBE}
\end{eqnarray}
%or in matrix form,
%\begin{eqnarray}
%&&\sum_{i',j'}R(u-v)^{i'j'}_{ij}T(u)^{i''}_{i'}T(v)^{j''}_{j'}(-1)^{[i'']([j']+[j''])}
%    \nonumber\\ && \mbox{} \quad\quad
%=\sum_{i',j'}T(v)^{j'}_{j}T(u)^{i'}_{i}R(u-v)^{i''j''}_{i'j'}(-1)^{[i]([j]+[j'])}.
%\end{eqnarray}

Define the transfer matrix $t(u)$
\begin{eqnarray}
t(u)=str_0T(u),\label{de:t}
\end{eqnarray}
where $str_0$ denotes the supertrace over the auxiliary space.
Then the Hamiltonian of the supersymmetric $t$-$J$ model is given
by
\begin{equation}
H={d\ln t(u)\over du}|_{u=0}. \label{de:H}
\end{equation}
This model is integrable thanks to the commutativity of the
transfer matrix for different parameters,
\begin{equation}
[t(u),t(v)]=0,
\end{equation}
which can be verified by using the GYBE.

Following \cite{Maillet96}, we now introduce the notation
$R^\sigma_{1\ldots N}$, where $\sigma$ is any element of the
permutation group ${\cal S}_N$. We note that we may rewrite the
GYBE as
\begin{eqnarray}
R_{23}^{\sigma_{23}}T_{0,23}=T_{0,32}R_{23}^{\sigma_{23}},
    \label{eq:RT-sigma23}
\end{eqnarray}
where $T_{0,23}\equiv R_{03}R_{02}$ and $\sigma_{23}$ is the
transposition of space labels (2,3). It follows that $R_{1\ldots
N}^{\sigma}$ is a product of elementary $R$-matrices,
corresponding to a decomposition of $\sigma$ into elementary
transpositions. With the help of the GYBE, one may generalize
(\ref{eq:RT-sigma23}) to a $N$-fold tensor product of spaces
\begin{eqnarray}
 R_{1\ldots N}^{\sigma}T_{0,1\ldots N}
  =T_{0,\sigma(1\ldots N)}R_{1\ldots N}^{\sigma},
     \label{eq:RT-sigma}
\end{eqnarray}
where $T_{0,1\ldots N}\equiv R_{0N}\ldots R_{01}.$ This implies
the ``decomposition" law
\begin{eqnarray}
 R^{\sigma'\sigma}_{1\ldots N}
 =R^{\sigma}_{\sigma'(1\ldots N)}
  R^{\sigma'}_{1\ldots N},\label{eq:R-RR}
\end{eqnarray}
for a product of two elements in ${\cal S}_N$. Note that
$R^{\sigma}_{\sigma'(1\ldots N)}$ satisfies the relation
\begin{eqnarray}
 R^{\sigma}_{\sigma'(1\ldots N)}
 T_{0,\sigma'(1\ldots N)}
 =T_{0,\sigma'\sigma(1\ldots N)}
  R^{\sigma}_{\sigma'(1\ldots N)}. \label{eq:RT-sigma'}
\end{eqnarray}
As in \cite{Albert00}, we write the elements of $R_{1\ldots
N}^{\sigma}$ as
\begin{eqnarray}
 \left(R_{1\ldots N}^{\sigma}\right)
  ^{\alpha_{\sigma(N)}\ldots \alpha_{\sigma(1)}}
  _{\beta_N\ldots \beta_1},
\end{eqnarray}
where the labels in the upper indices are permuted relative to the
lower indices according to $\sigma$.

\sect{$F$-matrices for the supersymmetric $t$-$J$ model}
 ~~~
In \cite{Maillet96}, Maillet and Sanchez de Santos constructed the
Drinfeld factorizing twists, i.e. the so-called factorizing
$F$-matrices, of the XXX model:
\begin{eqnarray}
R_{12}=F_{21}^{-1}F_{12}\ .\label{eq:R-F}
\end{eqnarray}
In \cite{Albert00}, Albert et al generalized the results in
\cite{Maillet96} to the $gl(m)$ spin chain system. In this
section, we construct the $F$-matrices associated with the
supersymmetric $t$-$J$ model.

\subsection{The $F$-matrix}
 ~~~~
 For the $R$-matrix (\ref{de:R}), we define the
$F$-matrix
\begin{eqnarray}
 &&F_{12}%\nonumber\\ &=&
         =\left(\begin{array}{ccccccccc}
 1&0&0 & 0&0&0 & 0&0&0\\
 0&1&0 & 0&0&0 & 0&0&0 \\
 0&0&1 & 0&0&0  & 0&0&0\\
 0&b_{12}&0 & a_{12}&0&0  & 0&0&0\\
 0&0&0      & 0&1&0  & 0&0&0\\
 0&0&0      & 0&0&1  & 0&0&0\\
 0&0&b_{12} & 0&0&0       & a_{12}&0&0\\
 0&0&0      & 0&0&b_{12}  & 0&a_{12}&0\\
 0&0&0      & 0&0&0       & 0&0&1+c_{12}
 \end{array} \right).% \nonumber\\
 \end{eqnarray}
It is convenient to write the $F$-matrix as the form
\begin{eqnarray}
F_{12}=\sum_{3\geq\alpha_2\geq\alpha_1}
      P_1^{\alpha_1}P_2^{\alpha_2}+c_{12}P_1^3P_2^3
      +\sum_{3\geq\alpha_1>\alpha_2}
      P_1^{\alpha_1}P_2^{\alpha_2}R_{12},\label{de:F12}
\end{eqnarray}
where $(P_i^\alpha)_{k}^{l}=\delta_{k,\alpha}\delta_{l,\alpha}$ is
the projector acting on $i$th space. Then by the $R$-matrix
(\ref{de:R}) and $F$-matrix (\ref{de:F12}), we have
\begin{eqnarray}
 F_{21}R_{12}%=\nonumber\\
&=&\left(\sum_{3\geq\alpha_1\geq\alpha_2}
      P_2^{\alpha_2}P_1^{\alpha_1}+c_{21}P_2^3P_1^3
      +\sum_{3\geq\alpha_2>\alpha_1}
      P_2^{\alpha_2}P_1^{\alpha_1}R_{21}\right)R_{12}
      \nonumber\\
&=&\left(\sum_{3\geq\alpha_1>\alpha_2}
      P_2^{\alpha_2}P_1^{\alpha_1}
      +P_2^1P_1^1+P_2^2P_1^2+(1+c_{21})P_2^3P_1^3
        \right. \nonumber\\&& \mbox{} \left. ~~~~~~~~~~~~
      +\sum_{3\geq\alpha_2>\alpha_1}
      P_2^{\alpha_2}P_1^{\alpha_1}R_{21}\right)R_{12}
      \nonumber\\
&=&\sum_{3\geq\alpha_1>\alpha_2}
      P_2^{\alpha_2}P_1^{\alpha_1}R_{12}
      +P_2^1P_1^1+P_2^2P_1^2+(1+c_{12})P_2^3P_1^3
      +\sum_{3\geq\alpha_2>\alpha_1}
      P_2^{\alpha_2}P_1^{\alpha_1}
      \nonumber\\
&=&\sum_{3\geq\alpha_1>\alpha_2}
      P_2^{\alpha_2}P_1^{\alpha_1}R_{12}
      +c_{12}P_2^3P_1^3
      +\sum_{3\geq\alpha_2\geq\alpha_1}
      P_2^{\alpha_2}P_1^{\alpha_1}
      \nonumber\\
&=&F_{12}.
\end{eqnarray}
Here we have used $R_{12}R_{21}=1$ and $c_{12}c_{21}=1$. Some
remarks are in order. The solutions to (\ref{eq:R-F}), i.e. the
$F$-matrices satisfying (\ref{eq:R-F}), are not unique
\cite{Maillet96,Albert00}. In this paper, we only consider the
particular solution (\ref{de:F12}), which is lower-triangle.

We now generalize the $F$-matrix to the $N$-site problem. As is
pointed out in \cite{Albert00}, the generalized $F$-matrix should
satisfy the three properties: i) lower-triangularity; ii)
non-degeneracy and
\begin{eqnarray}
\mbox{iii)}\ \ F_{\sigma(1\ldots
N)}(u_{\sigma(1)},\ldots,u_{\sigma(N)})
  R_{1\ldots N}^\sigma(u_1,\ldots,u_N)
 =F_{1\ldots N}(u_1,\ldots,u_N), \label{eq:R-F-N}
\end{eqnarray}
where $\sigma\in {\cal S}_N$ and $u_i$, $i=1,\ldots,N$, are
generic inhomogeneous parameters .

Define the $N$-site $F$-matrix:
\begin{eqnarray}
F_{1,\ldots N}=\sum_{\sigma\in {\cal S}_N}
   \sum_{\alpha_{\sigma(1)}\ldots\alpha_{\sigma(N)}}^{\quad\quad *}
   \prod_{j=1}^N P_{\sigma(j)}^{\alpha_{\sigma(j)}}
   S(c,\sigma,\alpha_\sigma)R_{1\ldots N}^\sigma, \label{de:F}
\end{eqnarray}
where the sum $\sum^*$  is over all non-decreasing sequences of
the labels $\alpha_{\sigma(i)}$:
\begin{eqnarray}
&& \alpha_{\sigma(i+1)}\geq \alpha_{\sigma(i)}\quad \mbox{if}\quad
              \sigma(i+1)>\sigma(i) \nonumber\\
&& \alpha_{\sigma(i+1)}> \alpha_{\sigma(i)}\quad \mbox{if}\quad
              \sigma(i+1)<\sigma(i) \label{cond:F}
\end{eqnarray}
and the c-number function $S(c,\sigma,\alpha_\sigma)$ is given by
\begin{eqnarray}
S(c,\sigma,\alpha_\sigma)\equiv
\exp\{\sum_{l>k=1}^N\delta^{[3]}_{\alpha_{\sigma(k)},\alpha_{\sigma(l)}}
    \ln(1+c_{\sigma(k)\sigma(l)})\}
\end{eqnarray}
with  $\delta^{[3]}_{\alpha_{\sigma(k)},\alpha_{\sigma(l)}}=1$ for
$\alpha_{\sigma(k)}=\alpha_{\sigma(l)}=3$, and
$\delta^{[3]}_{\alpha_{\sigma(k)},\alpha_{\sigma(l)}}=0$
otherwise.

 %Note that $R^{\sigma}$ is factorized according to $R$'s of
 % the elementary transpositions.
The definition of $F_{1\ldots N}$, (\ref{de:F}), and the summation
condition (\ref{cond:F}) imply that $F_{1\ldots N}$ is a
lower-triangular matrix. Moreover, one can easily check that the
$F$-matrix is non-degenerate because all diagonal elements are
non-zero.

We now prove that the $F$-matrix (\ref{de:F}) satisfies the
property iii). Any given permutation $\sigma\in {\cal S}_N$ can be
decomposed into elementary transpositions of the group ${\cal
S}_N$ as $\sigma=\sigma_1\ldots \sigma_k$ with $\sigma_i$ denoting
the elementary permutation $(i,i+1)$. By (\ref{eq:R-RR}), we have
if the property iii) holds for elementary transposition
$\sigma_i$,
\begin{eqnarray}
&&F_{\sigma(1\ldots N)}R^{\sigma}_{1\ldots N}= \nonumber\\
 &=& F_{\sigma_1\ldots\sigma_k(1\ldots N)}
     R^{\sigma_k}_{\sigma_1\ldots\sigma_{k-1}(1\ldots N)}
     R^{\sigma_{k-1}}_{\sigma_1\ldots\sigma_{k-2}(1\ldots N)}
     \ldots
     R^{\sigma_1}_{1\ldots N}\nonumber\\
 &=&F_{\sigma_1\ldots\sigma_{k-1}(1\ldots N)}
     R^{\sigma_{k-1}}_{\sigma_1\ldots\sigma_{k-2}(1\ldots N)}
     \ldots
     R^{\sigma_1}_{1\ldots N}\nonumber\\
  &=&\ldots
     =F_{\sigma_1(1\ldots N)}R^{\sigma_1}_{1\ldots N}=F_{1\ldots N}.
\end{eqnarray}

For the elementary transposition $\sigma_i$, we have
\begin{eqnarray}
 &&F_{\sigma_i(1\ldots N)}R^{\sigma_i}_{1\ldots N}= \nonumber\\
 &=&\sum_{\sigma\in {\cal S}_N}
   \sum_{\alpha_{\sigma_i\sigma(1)}\ldots\alpha_{\sigma_i\sigma(N)}}^{\quad\quad *}
   \prod_{j=1}^N P_{\sigma_i\sigma(j)}^{\alpha_{\sigma_i\sigma(j)}}
 %  \nonumber\\ &&\times
  S(c,\sigma_i\sigma,\alpha_{\sigma_i\sigma})R_{\sigma_i(1\ldots N)}^\sigma
   R_{1\ldots N}^{\sigma_i} \nonumber\\
 &=&\sum_{\sigma\in {\cal S}_N}
   \sum_{\alpha_{\sigma_i\sigma(1)}\ldots\alpha_{\sigma_i\sigma(N)}}^{\quad\quad *}
   \prod_{j=1}^N P_{\sigma_i\sigma(j)}^{\alpha_{\sigma_i\sigma(j)}}
%   \nonumber\\ &&\times
   S(c,\sigma_i\sigma,\alpha_{\sigma_i\sigma})
   R_{1\ldots N}^{\sigma_i\sigma} \nonumber\\
 &=&\sum_{\tilde\sigma\in {\cal S}_N}
   \sum_{\alpha_{\tilde\sigma(1)}\ldots\alpha_{\tilde\sigma(N)}}^{\quad\quad *(i)}
   \prod_{j=1}^N P_{\tilde\sigma(j)}^{\alpha_{\tilde\sigma(j)}}
   S(c,\tilde\sigma,\alpha_{\tilde\sigma})R_{1\ldots N}^{\tilde\sigma}, \label{eq:FR-F}\nonumber\\
\end{eqnarray}
where $\tilde\sigma=\sigma_i\sigma$, and the summation sequences
of $\alpha_{\tilde\sigma}$ in ${\sum^*}^{(i)}$ now has the form
\begin{eqnarray}
&& \alpha_{\tilde\sigma(j+1)}\geq \alpha_{\tilde\sigma(j)}\quad
\mbox{if}\quad
              \sigma_i\tilde\sigma(j+1)>\sigma_i\tilde\sigma(j), \nonumber\\
&& \alpha_{\tilde\sigma(j+1)}> \alpha_{\tilde\sigma(j)}\quad
\mbox{if}\quad
              \sigma_i\tilde\sigma(j+1)<\sigma_i\tilde\sigma(j). \label{cond:FR-F}
\end{eqnarray}
Comparing (\ref{cond:FR-F}) with (\ref{cond:F}), we find that the
only difference between them is the transposition $\sigma_i$
factor in the ``if" conditions.  For a given $\tilde\sigma\in
{\cal S}_N$ with $\tilde\sigma(j)=i$ and $\tilde\sigma(k)=i+1$, we
now examine how the elementary transposition $\sigma_i$ will
affect the inequalities (\ref{cond:FR-F}).\\
1). If $|j-k|>1$, then $\sigma_i$ does not affect the sequence of
$\alpha_{\tilde\sigma}$ at all, that is, the sign of inequality
$``>"$ or ``$\geq$" between two neighboring root indexes is
unchanged
with the action of $\sigma_i$. \\
2). If $|j-k|=1$, then in the summation sequences of
$\alpha_{\tilde\sigma}$, when $\tilde\sigma(j+1)=i+1$ and
$\tilde\sigma(j)=i$, sign ``$\geq$"  changes to $``>"$, while when
$\tilde\sigma(j+1)=i$ and $\tilde\sigma(j)=i+1$, $``>"$ changes to
``$\geq$". Thus (\ref{cond:F}) and (\ref{eq:FR-F}) differ only
when equal labels $\alpha_{\tilde\sigma}$ appear. In the
following, we study this difference between (\ref{de:F}) and
(\ref{eq:FR-F}). Rewriting $\tilde\sigma$ as $\sigma$, and then
subtracting $F_{1\ldots N}$ from $F_{\sigma_i(1\ldots
N)}R^{\sigma_i}_{1\ldots N}$, we have
\begin{eqnarray}
 &&F_{\sigma_i(1\ldots N)}R^{\sigma_i}_{1\ldots N}- F_{1\ldots N}=
      \nonumber\\
  &=&\sum_{\stackrel{\sigma(j)=i+1}{\sigma(j+1)=i}}
     \sum_{\alpha_{\sigma(j)}=\alpha_{\sigma(j+1)}=1}^3
     \sum_{\alpha_{\sigma(1)}\ldots\alpha_{\sigma(j-1)}
           \alpha_{\sigma(j+2)}\ldots\alpha_{\sigma(N)}}^{\quad\quad*}
     P_{\sigma(1)}^{\alpha_{\sigma(1)}}\ldots
      P_{i+1}^{\alpha_{i+1}}P_{i}^{\alpha_{i}}\ldots
      P_{\sigma(N)}^{\alpha_{\sigma(N)}}\nonumber\\
  && \quad\quad\quad\quad\quad \times
     S(c,\sigma,\alpha_\sigma)R_{1\ldots N}^{\sigma} \nonumber\\
  &&-\ \sum_{\stackrel{\sigma(j)=i}{\sigma(j+1)=i+1}}
      \sum_{\alpha_{\sigma(j)}=\alpha_{\sigma(j+1)}=1}^3
      \sum_{\alpha_{\sigma(1)}\ldots\alpha_{\sigma(j-1)}
           \alpha_{\sigma(j+2)}\ldots\alpha_{\sigma(N)}}^{\quad\quad*}
     P_{\sigma(1)}^{\alpha_{\sigma(1)}}\ldots
      P_{i}^{\alpha_i}P_{i+1}^{\alpha_{i+1}}\ldots
      P_{\sigma(N)}^{\alpha_{\sigma(N)}}\nonumber\\
  && \quad\quad\quad\quad\quad \times
     S(c,\sigma,\alpha_\sigma)R_{1\ldots N}^{\sigma}
   \label{eq:F-F1}\\
&=&\sum_{\stackrel{\sigma(j)=i+1}{\sigma(j+1)=i}}
    \sum_{\alpha_{i+1}=\alpha_{i}=1}^3
    \sum_{\alpha_{\sigma(1)}\ldots\alpha_{\sigma(j-1)}
           \alpha_{\sigma(j+2)}\ldots\alpha_{\sigma(N)}}^{\quad\quad*}
      P_{\sigma(1)}^{\alpha_{\sigma(1)}}\ldots
      P_{i+1}^{\alpha_{i+1}}P_{i}^{\alpha_i}\ldots
      P_{\sigma(N)}^{\alpha_{\sigma(N)}}
       \nonumber\\ &&\times
      \exp\{\sum_{\stackrel{1=k<l}{k,l\ne j,j+1}}^N
       \delta^{[3]}_{\alpha_{\sigma(k)},\alpha_{\sigma(l)}}
    \ln(1+c_{\sigma(k),\sigma(l)})\}
    \exp\{\delta^{[3]}_{\alpha_{i+1},\alpha_{i}}
     \ln(1+c_{i+1,i})\}
   R_{1\ldots N}^{\sigma} \nonumber\\
  &&\ -\sum_{\stackrel{\sigma(j)=i}{\sigma(j+1)=i+1}}
     \sum_{\alpha_{i}=\alpha_{i+1}=1}^3
     \sum_{\alpha_{\sigma(1)}\ldots\alpha_{\sigma(j-1)}
           \alpha_{\sigma(j+2)}\ldots\alpha_{\sigma(N)}}^{\quad\quad*}
      P_{\sigma(1)}^{\alpha_{\sigma(1)}}\ldots
      P_{i}^{\alpha_{i}}P_{i+1}^{\alpha_{i+1}}\ldots
      P_{\sigma(N)}^{\alpha_{\sigma(N)}}
       \nonumber\\ &&\times
      \exp\{\sum_{\stackrel{1=k<l}{k,l\ne j,j+1}}^N
      \delta^{[3]}_{\alpha_{\sigma(k)},\alpha_{\sigma(l)}}
    \ln(1+c_{\sigma(k),\sigma(l)})\}
     \exp\{\delta^{[3]}_{\alpha_{i},\alpha_{i+1}}
     \ln(1+c_{i,i+1})\}
   R_{1\ldots N}^{\sigma}.  \label{eq:F-F2} \nonumber\\
\end{eqnarray}
Making the change $\sigma\rightarrow\sigma\sigma_j$ in the first
term of the r.h.s. and using (\ref{eq:R-RR}), we have
\begin{eqnarray}
&&F_{\sigma_i(1\ldots N)}R^{\sigma_i}_{1\ldots N}- F_{1\ldots N}=
      \nonumber\\
 &=&\sum_{\stackrel{\sigma(j)=i}{\sigma(j+1)=i+1}}
    \sum_{\alpha_{\sigma(1)}\ldots\alpha_{\sigma(j-1)}
           \alpha_{\sigma(j+2)}\ldots\alpha_{\sigma(N)}}^{\quad\quad*}
     P_{\sigma(1)}^{\alpha_{\sigma(1)}}\ldots P_{\sigma(i-1)}^{\alpha_{\sigma(i-1)}}
     P_{\sigma(i+2)}^{\alpha_{\sigma(i+2)}}\ldots P_{\sigma(N)}^{\alpha_{\sigma(N)}}
     \nonumber\\ &&\times
     \exp\{\sum_{\stackrel{1=k<l}{k,l\ne j,j+1}}^N
       \delta^{[3]}_{\alpha_{\sigma(k)},\alpha_{\sigma(l)}}
    \ln(1+c_{\sigma(k),\sigma(l)})\}\nonumber\\
&&\times \left[
    \sum_{\alpha_{i+1}=\alpha_{i}=1}^3\left(
      P_{i+1}^{\alpha_{i+1}}P_{i}^{\alpha_i}
    \exp\{\delta^{[3]}_{\alpha_{i+1},\alpha_{i}}
     \ln(1+c_{i+1,i})\}R_{\sigma(1\ldots N)}^{\sigma_j}
      \right.\right.
    \nonumber\\ && \mbox{}\left.\left.
   - P_{i}^{\alpha_{i}}P_{i+1}^{\alpha_{i+1}}
    \exp\{\delta^{[3]}_{\alpha_{i},\alpha_{i+1}}
     \ln(1+c_{i,i+1})\}\right)\right]R_{1\ldots N}^{\sigma}
\end{eqnarray}
Denoted by $X$ the quantity in the square bracket. Then we have
\begin{eqnarray}
X&=&\left(P_{i+1}^1 P_{i}^1+P_{i+1}^2 P_{i}^2
    +(1+c_{i+1,i})P_{i+1}^3 P_{i}^3\right)
    R^{\sigma_j}_{\sigma(1\ldots N)} \nonumber \\
  &&\mbox{}  - \left(P_{i}^1 P_{i+1}^1+P_{i}^2 P_{i+1}^2
    +(1+c_{i,i+1})P_{i}^3 P_{i+1}^3 \right)\nonumber\\
 &=&\left(P_{i+1}^1 P_{i}^1+P_{i+1}^2 P_{i}^2
    +(1+c_{i+1,i})P_{i+1}^3 P_{i}^3\right)
    R_{i,i+1} \nonumber \\
  &&\mbox{}  - \left(P_{i}^1 P_{i+1}^1+P_{i}^2 P_{i+1}^2
    +(1+c_{i,i+1})P_{i}^3 P_{i+1}^3\right) \nonumber\\
 &=&0. \label{X}
\end{eqnarray}
Thus, we obtain
\begin{eqnarray}
R_{1\ldots N}^\sigma(u_1,\ldots,u_N)
 =F^{-1}_{\sigma(1\ldots N)}(u_{\sigma(1)},\ldots,
   u_{\sigma(N)})F_{1\ldots N}(u_1,\ldots,u_N).
\end{eqnarray}
The factorizing $F$-matrix $F_{1\ldots N}$ of the supersymmetric
$t$-$J$ model is proved to satisfy all three properties.%\\[3mm]

\subsection{The inverse $F^{-1}_{1\ldots N}$ of the $F$-matrix }
 ~~~
 The non-degenerate property of the $F$-matrix implies that we can find
the inverse matrix $F^{-1}_{1\ldots N}$. To do so, we first define
the $F^*$-matrix
\begin{eqnarray}
F^*_{1\ldots N}&=&\sum_{\sigma\in {\cal S}_N}
   \sum_{\alpha_{\sigma(1)}\ldots\alpha_{\sigma(N)}}^{\quad\quad **}
   S(c,\sigma,\alpha_\sigma)R_{\sigma(1\ldots N)}^{\sigma^{-1}}
   \prod_{j=1}^N P_{\sigma(j)}^{\alpha_{\sigma(j)}},
    \label{de:F*}  \nonumber\\
\end{eqnarray}
where the sum $\sum^{**}$ is taken over all possible $\alpha_i$
which satisfies the following non-increasing constraints:
\begin{eqnarray}
&& \alpha_{\sigma(i+1)}\leq \alpha_{\sigma(i)}\quad \mbox{if}\quad
              \sigma(i+1)<\sigma(i), \nonumber\\
&& \alpha_{\sigma(i+1)}< \alpha_{\sigma(i)}\quad \mbox{if}\quad
              \sigma(i+1)>\sigma(i). \label{cond:F*}
\end{eqnarray}
%and $S^*(c,\sigma,\alpha_\sigma)$ is given by
%\begin{eqnarray}
%S^*(c,\sigma,\alpha_\sigma)=\exp\{\sum_{k>l=1}^N\delta^{[3]}_{\alpha_{\sigma(k)},\alpha_{\sigma(l)}}
%    \ln(1+c_{\sigma(k)\sigma(l)})\}
%\end{eqnarray}

Now we compute the product of $F_{1\ldots N}$ and $F^*_{1\ldots
N}$. Substituting (\ref{de:F}) and (\ref{de:F*}) into the product,
we have
\begin{eqnarray}
 F_{1\ldots N}F^*_{1\ldots N}% \nonumber\\
 &=&\sum_{\sigma\in {\cal S}_N}\sum_{\sigma'\in {\cal S}_N}
    \sum^{\quad\quad *}_{\alpha_{\sigma_1}\ldots\alpha_{\sigma_N}}
    \sum^{\quad\quad **}_{\beta_{\sigma'_1}\ldots\beta_{\sigma'_{N}}}
    S(c,\sigma,\alpha_\sigma)S(c,\sigma',\beta_{\sigma'})
    \nonumber\\ &&\times
    \prod_{i=1}^N P_{\sigma(i)}^{\alpha_{\sigma(i)}}
    R^{\sigma}_{1\ldots N}R^{{\sigma'}^{-1}}_{\sigma'(1\ldots N)}
    \prod_{i=1}^N P_{\sigma'(i)}^{\beta_{\sigma'(i)}} \nonumber\\
 &=&\sum_{\sigma\in {\cal S}_N}\sum_{\sigma'\in {\cal S}_N}
    \sum^{\quad\quad *}_{\alpha_{\sigma_1}\ldots\alpha_{\sigma_N}}
    \sum^{\quad\quad **}_{\beta_{\sigma'_1}\ldots\beta_{\sigma'_{N}}}
    S(c,\sigma,\alpha_\sigma)S(c,\sigma',\beta_{\sigma'})
     \nonumber\\ &&\times
    \prod_{i=1}^N P_{\sigma(i)}^{\alpha_{\sigma(i)}}
    R^{{\sigma'}^{-1}\sigma}_{\sigma'(1\ldots N)}
    \prod_{i=1}^N P_{\sigma'(i)}^{\beta_{\sigma'(i)}}. \label{eq:FF*-1}
\end{eqnarray}
To evaluate the r.h.s., we examine the matrix element of the
$R$-matrix
\begin{eqnarray}
\left(R^{{\sigma'}^{-1}\sigma}_{\sigma'(1\ldots N)}\right)
  ^{\alpha_{\sigma(N)}\ldots\alpha_{\sigma(1)}}
  _{\beta_{\sigma'(N)}\ldots\beta_{\sigma'(1)}}.
  \label{eq:R-index}
\end{eqnarray}
Note that the sequence $\{ \alpha_{\sigma}\}$ is non-decreasing
and $\{\beta_{\sigma'}\}$ is non-increasing. Thus the
non-vanishing condition of the matrix element (\ref{eq:R-index})
requires that $\alpha_{\sigma}$ and $\beta_{\sigma'}$ satisfy
\begin{eqnarray}
\beta_{\sigma'(N)}=\alpha_{\sigma(1)},\ldots,
\beta_{\sigma'(1)}=\alpha_{\sigma(N)}. \label{re:alpha-beta}
\end{eqnarray}
One can verify \cite{Albert00} that (\ref{re:alpha-beta}) is
fulfilled only if
\begin{eqnarray}
\sigma'(N)=\sigma(1),\ldots,\sigma'(1)=\sigma(N).
\label{re:sigma-sigma'}
\end{eqnarray}

Let $\bar\sigma$ be the maximal element of the ${\cal S}_N$ which
reverses the site labels
\begin{eqnarray}
\bar\sigma(1,\ldots,N)=(N,\ldots,1).
\end{eqnarray}
Then from (\ref{re:sigma-sigma'}), we have
\begin{eqnarray}
\sigma'=\sigma\bar\sigma. \label{eq:sigma'}
\end{eqnarray}
Substituting (\ref{re:alpha-beta}) and (\ref{eq:sigma'}) into
(\ref{eq:FF*-1}), we have
\begin{eqnarray}
 F_{1\ldots N}F^*_{1\ldots N}%= \nonumber\\
 &=&\sum_{\sigma\in {\cal S}_N}
    \sum^{\quad\quad *}_{\alpha_{\sigma_1}\ldots\alpha_{\sigma_N}}
    S(c,\sigma,\alpha_\sigma)S(c,\sigma,\alpha_\sigma)
    \prod_{i=1}^N P_{\sigma(i)}^{\alpha_{\sigma(i)}}
    R^{\bar\sigma}_{\sigma(N\ldots 1)}
    \prod_{i=1}^N P_{\sigma(i)}^{\alpha_{\sigma(i)}}\label{eq:FF*-2}.
      \nonumber\\
\end{eqnarray}
The decomposition of $R^{\bar\sigma}$ in terms of elementary
$R$-matrices is unique module GYBE. One reduces from
(\ref{eq:FF*-2}) that $FF^*$ is a diagonal matrix:
\begin{eqnarray}
F_{1\ldots N}F^*_{1\ldots N}=\prod_{i<j}\Delta_{ij},
\end{eqnarray}
where
\begin{eqnarray}
[\Delta_{ij}]^{\beta_i\beta_j}_{\alpha_i\alpha_j} =
 \delta_{\alpha_i\beta_i}\delta_{\alpha_j\beta_j}\left\{
 \begin{array}{cl}
  a_{ij}& \mbox{if} \ \alpha_i>\alpha_j\\
  a_{ji}& \mbox{if} \ \alpha_i<\alpha_j\\
  1& \mbox{if}\ \alpha_i=\alpha_j=1,2 \\
  (1+c_{ij})(1+c_{ji})& \mbox{if}\ \alpha_i=\alpha_j=3
  \end{array}\right.. \label{de:F-Inv}
\end{eqnarray}

Therefore, the inverse of the $F$-matrix is given by
\begin{equation}
F^{-1}_{1\ldots N}=F^*_{1\ldots N}\prod_{i<j}\Delta_{ij}^{-1}.
\end{equation}

\sect{The monodromy matrix in the $F$-basis} ~~~ In the previous
section, we see that the $gl(2|1)$ $R$-matrix factorizes in terms
of the $F$-matrix and its inverse which we constructed explicitly.
The column vectors of the inverse of the $F$-matrix form a set of
basis on which $gl(2|1)$ acts. In this section, we study the
generators of $gl(2|1)$ and the elements of the monodromy matrix
in the $F$-basis.

\subsection{$gl(2|1)$ generators in the $F$-basis}
~~~ The $N$-site supersymmetric $gl(2|1)$ system has 4 simple
generators: $E^{12},E^{21},E^{23}$ and $E^{32}$ with $E^{\gamma\
\gamma\pm 1}=E^{\gamma\,\gamma\pm 1}_{(1)} +\ldots+E^{\gamma\
\gamma\pm 1}_{(N)}$, where $E^{\gamma\,\gamma\pm 1}_{(k)}$ acts on
the $k$th component of the tensor product space. Let $\tilde
E^{\gamma\,\gamma\pm 1}$ denote the corresponding simple
generators in the $F$-basis: $\tilde E^{\gamma\, \gamma\pm
1}=F_{1\ldots N}E^{\gamma\, \gamma\pm 1}F_{1\ldots N}^{-1}$. We
first derive the $\tilde E^{12}$. From the expressions of $F$ and
its inverse, we have
\begin{eqnarray}
 \tilde E^{12}%\nonumber\\
 &=&F_{1\ldots N}E^{12}F^{-1}_{1\ldots N}\nonumber\\
 &=&\sum_{\sigma,\sigma'\in {\cal S}_N}
    \sum^{\quad\quad*}_{\alpha_{\sigma(1)}\ldots\alpha_{\sigma(N)}}
    \sum^{\quad\quad**}_{\beta_{\sigma'(1)}\ldots\beta_{\sigma'(N)}}
    S(c,\sigma,\alpha_\sigma)S(c,\sigma',\beta_{\sigma'})
    \nonumber\\ && \times
    \prod_{i=1}^N P_{\sigma(i)}^{\alpha_{\sigma(i)}}
    R^{\sigma}_{1\ldots N}E^{12}R^{{\sigma'}^{-1}}_{\sigma'(1\ldots N)}
    \prod_{i=1}^N
    P_{\sigma'(i)}^{\beta_{\sigma'(i)}}\prod_{i<j}\Delta_{ij}^{-1}
     \nonumber\\
 &=&\sum_{\sigma,\sigma'\in {\cal S}_N}
    \sum^{\quad\quad*}_{\alpha_{\sigma(1)}\ldots\alpha_{\sigma(N)}}
    \sum^{\quad\quad**}_{\beta_{\sigma'(1)}\ldots\beta_{\sigma'(N)}}
    S(c,\sigma,\alpha_\sigma)S(c,\sigma',\beta_{\sigma'})
    \nonumber\\ &&\times
    \prod_{i=1}^N P_{\sigma(i)}^{\alpha_{\sigma(i)}}E^{12}
    R^{{\sigma'}^{-1}\sigma}_{\sigma'(1\ldots N)}
    \prod_{i=1}^N P_{\sigma'(i)}^{\beta_{\sigma'(i)}}
    \prod_{i<j}\Delta_{ij}^{-1}
    \label{eq:E-1}     \\
 &=&\sum_{\sigma,\sigma'\in {\cal S}_N}
    \sum_{k=1}^N E^{12}_{(\sigma(l))}
    \sum^{\quad\quad*}_{\alpha_{\sigma(1)}\ldots\alpha_{\sigma(N)}}
    \sum^{\quad\quad**}_{\beta_{\sigma'(1)}\ldots\beta_{\sigma'(N)}}
    S(c,\sigma,\alpha_\sigma)S(c,\sigma',\beta_{\sigma'}) \nonumber\\ && \times
    P_{\sigma(1)}^{\alpha_{\sigma(1)}=1}\ldots
    \left(P_{\sigma(l)=k}^{\alpha_{\sigma(l)}=1\rightarrow 2}\right)
    \ldots P_{\sigma(N)}^{\alpha_{\sigma(N)}}
    R^{{\sigma'}^{-1}\sigma}_{\sigma'(1\ldots N)}
    \prod_{i=1}^N
    P_{\sigma'(i)}^{\beta_{\sigma'(i)}}\prod_{i<j}\Delta_{ij}^{-1},
    \label{eq:E-2} \nonumber\\
\end{eqnarray}
where in (\ref{eq:E-1}), we have used $[E^{\gamma\,\gamma\pm
1},R^{\sigma}_{1\ldots N}]=0.$ The element of
$R^{{\sigma'}^{-1}\sigma}_{\sigma'(1\ldots N)}$ between $
P_{\sigma(1)}^{\alpha_{\sigma(1)}=1}\ldots
    \left(P_{\sigma(l)=k}^{\alpha_{\sigma(l)}=1\rightarrow 2}\right)
    \ldots P_{\sigma(N)}^{\alpha_{\sigma(N)}}$ and $
 P_{\sigma'(N)}^{\beta_{\sigma'(N)}}\ldots
   P_{\sigma'(1)}^{\beta_{\sigma'(1)}}$
is denoted as
 \begin{eqnarray}
 \left(R^{{\sigma'}^{-1}\sigma}_{\sigma'(1\ldots N)}\right)
 ^{\stackrel{\sigma(N)}{\alpha_{\sigma(N)}}\ldots
 \stackrel{\sigma(l)=k}{1\rightarrow 2}
  \ldots \stackrel{\sigma(1)}{1}}
 _{\beta_{\sigma'(N)}\ldots \beta_{\sigma'(1)}}. \label{eq:E-R-index}
 \end{eqnarray}
We call the sequence $\{\alpha_{\sigma(l)}\}$ {\bf normal} if it
is arranged according to the rules in (\ref{cond:F}), otherwise,
we call it {\bf abnormal}.

It is now convenient for us to discuss the non-vanishing condition
of the $R$-matrix element (\ref{eq:E-R-index}).  Comparing
(\ref{eq:E-R-index}) with (\ref{eq:R-index}), we find that the
difference between them lies in the $k$th site. Because the group
label in the $k$th space has been changed, the sequence
$\{\alpha_\sigma\}$ is now a abnormal sequence. However, it can be
permuted to the normal sequence by some permutation $\hat
\sigma_k$. Namely, $\alpha_{1\rightarrow 2}$ in the abnormal
sequence  can be moved to a suitable position by using the
permutation $\hat \sigma_k$ according to rules in (\ref{cond:F}).
(It is easy to verify that $\hat\sigma_k$ is unique by using
(\ref{cond:F}).) Thus, by procedure similar to that in the
previous section, we find that when
\begin{equation}
\sigma'=\hat\sigma_k\sigma\bar\sigma\quad \mbox{and}\quad
 \beta_{\sigma'(N)}=\alpha_{\sigma(1)},\ldots,
 \beta_{\sigma'(1)}=\alpha_{\sigma(N)},\label{eq:E-sigma}
\end{equation}
the $R$-matrix element (\ref{eq:E-R-index}) is non-vanishing.
% (We note that for
%a $\{P_{\sigma}^{\alpha_\sigma}\}$ sequence
%(\ref{eq:sqc-alpha}), we can find $m-l+1$
%$\{P_{\sigma'}^{\beta_\sigma'}\}$ sequences which ensure the
%non-vanishing of $R$-matrix according to the (\ref{cond:F}) and
%(\ref{cond:F*})).

Because the non-zero condition of the elementary $R$-matrix
element $R^{i'j'}_{ij}$ is $i+j=i'+j'$, the following $R$-matrix
elements
\begin{eqnarray}
 \left(R^{{\sigma'}^{-1}\sigma}_{\sigma'(1\ldots N)}\right)
 ^{\stackrel{\sigma(N)}{\alpha_{\sigma(N)}}\ldots
 \stackrel{\sigma(l)=k}{1} \ldots
 \stackrel{\sigma(n)}{1\rightarrow 2}
  \ldots \stackrel{\sigma(1)}{1}}
 _{\beta_{\sigma'(N)}\ldots \beta_{\sigma'(1)}} \label{eq:E-R-nindex}
 \end{eqnarray}
with $1\le n<l$ are also non-vanishing.

Therefore, (\ref{eq:E-2}) becomes
\begin{eqnarray}
 \tilde E^{12}%\nonumber\\
 &=&\sum_{\sigma\in {\cal S}_N}\sum_{k=1}^N
    \sum^{\quad\quad *}_{\alpha_{\sigma_1}\ldots\alpha_{\sigma_N}}
    S(c,\sigma,\alpha_\sigma)S(c,\hat\sigma_k\sigma,\alpha_{\hat\sigma_k\sigma})
       \nonumber\\ && \times
    \left[E^{12}_{(\sigma(l))}P_{\sigma(1)}^{\alpha_{\sigma(1)}=1}
    \ldots P_{\sigma(l)=k}^{\alpha_{\sigma(l)}=1\rightarrow 2}
    \ldots P_{\sigma(N)}^{\alpha_{\sigma(N)}}+\ldots+\right.
       \nonumber\\ && \mbox{} \quad
   +E^{12}_{(\sigma(n))}P_{\sigma(1)}^{\alpha_{\sigma(1)}=1}
    \ldots P_{\sigma(n)}^{\alpha_{\sigma(n)}=1\rightarrow 2}
    \ldots P_{\sigma(l)=k}^{\alpha_{\sigma(l)}=1}\ldots
    P_{\sigma(N)}^{\alpha_{\sigma(N)}}+ \ldots+
       \nonumber\\ && \mbox{}  \mbox{}\quad\left.
   +E^{12}_{(\sigma(1))}P_{\sigma(1)}^{\alpha_{\sigma(1)}=1\rightarrow 2}
    \ldots P_{\sigma(l)=k}^{\alpha_{\sigma(l)}=1}\ldots
    P_{\sigma(N)}^{\alpha_{\sigma(N)}}\right]
        \nonumber\\ && \times
    R^{\bar\sigma\sigma^{-1}\hat\sigma^{-1}_k\sigma}
     _{\hat\sigma_k\sigma(N\ldots 1)}
    \prod_{i=1}^N
    P_{\hat\sigma_k\sigma(i)}^{\alpha_{\hat\sigma_k\sigma(i)}}
    \prod_{i<j}\Delta_{ij}^{-1}, \label{eq:E-3}\\
 &=&\sum_{k=1}^N E^{12}_{(k)}\otimes_{j\ne k} G^{12}_{(j)}(k,j),
     \label{eq:E-12-tilde}
\end{eqnarray}
where $\hat\sigma_k$ is the element of  ${\cal S}_N$ which
permutes the first abnormal sequence in the square bracket of
(\ref{eq:E-3}) to normal sequence, and
\begin{equation}
  G^{12}(k,j)= \mbox{diag}(a_{kj}^{-1},1,1).
\end{equation}

Similarly, we obtain the expressions of other three simple
generators in the $F$-basis:
\begin{eqnarray}
\tilde E^{\gamma\,\gamma\pm 1}=\sum_{i=1}^N E^{\gamma\,\gamma\pm
1}_{(i)}\otimes_{j\ne i}
  G^{\gamma,\gamma\pm 1}_{(j)}(i,j) \label{eq:E-gamma-tilde}
\end{eqnarray}
with
\begin{eqnarray}
G^{21}(i,j)&=&\mbox{diag}(1,a_{ji}^{-1},1),\nonumber \\
G^{23}(i,j)&=&\mbox{diag}\left(1,a_{ij}^{-1},(2a_{ij})^{-1}\right),\nonumber\\
G^{32}(i,j)&=&\mbox{diag}(1,1,2).
\end{eqnarray}

With the help of the simple generators, the non-simple generators
$\tilde E^{13}$ and $\tilde E^{31}$ of $gl(2|1)$ can be obtained
by the commutation relations,
\begin{eqnarray}
 \tilde E^{13}=[\tilde E^{12},\tilde E^{23}],\quad
 \tilde E^{31}=[\tilde E^{32},\tilde E^{21}]. \label{eq:E-1331}
\end{eqnarray}
Substituting (\ref{eq:E-12-tilde}) and(\ref{eq:E-gamma-tilde})
 into (\ref{eq:E-1331}), we obtain
\begin{eqnarray}
\tilde E^{13}&=&\sum_{i=1}^N E^{13}_{(i)}\otimes_{j\ne i}
    \mbox{diag}\left(a_{ij}^{-1},a_{ij}^{-1},(2a_{ij})^{-1}\right)_{(j)}
           \nonumber\\ && \mbox{}
    +\sum_{i\ne j=1}^N {\eta\over z_j-z_i}
       E^{12}_{(i)}\otimes E^{23}_{(j)}\otimes_{k\ne i,j}
     \mbox{diag}\left(a_{ik}^{-1},a_{jk}^{-1},(2a_{jk})^{-1}\right)_{(k)},
           \nonumber\\
\tilde E^{31}&=&\sum_{i=1}^N E^{31}_{(i)}\otimes_{j\ne i}
    \mbox{diag}\left(1,a_{ji}^{-1},2\right)_{(j)}
           \nonumber\\ && \mbox{}
    +\sum_{i\ne j=1}^N {\eta\over z_i-z_j}
       E^{32}_{(i)}\otimes E^{21}_{(j)}\otimes_{k\ne i,j}
     \mbox{diag}\left(1,a_{kj}^{-1},2\right)_{(k)}.
\end{eqnarray}

\subsection{Elements of the monodromy matrix in the $F$-basis}
~~~~In the $F$-basis, the monodromy matrix $T(u)$,
(\ref{de:T-marix}), becomes
\begin{eqnarray}
 \tilde{T}(u)\equiv F_{1\ldots N}T(u)F^{-1}_{1\ldots N}
 =\left(\begin{array}{ccc}
\tilde{ A}_{11}(u)& \tilde{A}_{12}(u)&\tilde{ B}_1(u)\\
\tilde{ A}_{21}(u)& \tilde{ A}_{22}(u)&\tilde{ B}_2(u)\\
\tilde{ C}_{1}(u)& \tilde{ C}_{2}(u)&\tilde{ D}(u)
 \end{array}\right). \label{de:T-marix-F}
 \end{eqnarray}
We first study the diagonal element $\tilde D(u)$. Acting the
$F$-matrix on $D(u)$, we have
\begin{eqnarray}
 F_{1\ldots N}T^{33}
 &=&\sum_{\sigma\in {\cal S}_N}
    \sum_{\alpha_{\sigma(1)}\ldots\alpha_{\sigma(N)}}^{\quad\quad*}
    S(c,\sigma,\alpha_\sigma)\prod_{i=1}^N
    P_{\sigma(i)}^{\alpha_{\sigma}}R^\sigma_{1\ldots N}
    P_0^3 T_{0,1\ldots N}P_0^3 \nonumber\\
 &=&\sum_{\sigma\in {\cal S}_N}
    \sum_{\alpha_{\sigma(1)}\ldots\alpha_{\sigma(N)}}^{\quad\quad*}
    S(c,\sigma,\alpha_\sigma)\prod_{i=1}^N
    P_{\sigma(i)}^{\alpha_{\sigma}}
    P_0^3 T_{0,\sigma(1\ldots N)}P_0^3
    R^\sigma_{1\ldots N}.
\end{eqnarray}
Following \cite{Albert00}, we can split the sum $\sum^*$ according
to the number of the occurrences of the index 3.
\begin{eqnarray}
F_{1\ldots N}T^{33}
 &=&\sum_{\sigma\in {\cal S}_N}
    \sum_{k=0}^N
    \sum_{\alpha_{\sigma(1)}\ldots\alpha_{\sigma(N)}}^{\quad\quad*}
    S(c,\sigma,\alpha_\sigma)
    \prod_{j=N-k+1}^N \delta_{\alpha_{\sigma(j)},3}
    P_{\sigma(j)}^{\alpha_{\sigma(j)}} \nonumber\\ &&\times
    \prod_{j=1}^{N-k}P_{\sigma(j)}^{\alpha_{\sigma(j)}}
    P_0^3 T_{0,\sigma(1\ldots N)}P_0^3
    R^\sigma_{1\ldots N}. \label{eq:T-tilde-1}
\end{eqnarray}
Consider the prefactor of $R^\sigma_{1\ldots N}$. We have
\begin{eqnarray}
 &&\prod_{j=1}^{N-k}P_{\sigma(j)}^{\alpha_{\sigma(j)}}
   \prod_{j=N-k+1}^N
    P_{\sigma(j)}^{3}
    P_0^3 T_{0,\sigma(1\ldots N)}P_0^3\nonumber\\
 &=&\prod_{j=1}^{N-k}P_{\sigma(j)}^{\alpha_{\sigma(j)}}
   \prod_{j=N-k+1}^N\left(R_{0\,\sigma(j)}\right)^{33}_{33}
   P_0^3 T_{0,\sigma(1\ldots N-k)}P_0^3
   \prod_{j=N-k+1}^N P_{\sigma(j)}^{3} \nonumber\\
 &=&\prod_{i=N-k+1}^N c_{0\,\sigma(i)}
   \prod_{j=1}^{N-k}P_{\sigma(j)}^{\alpha_{\sigma(j)}}
   P_0^3 T_{0,\sigma(1\ldots N-k)}P_0^3
   \prod_{j=N-k+1}^N P_{\sigma(j)}^{3} \nonumber\\
 &=&\prod_{i=N-k+1}^N c_{0\,\sigma(i)}
   \prod_{i=1}^{N-k}\left(R_{0\,\sigma(i)}\right)
            ^{3\alpha_{\sigma(i)}}_{3\alpha_{\sigma(i)}}
   \prod_{j=1}^{N-k}P_{\sigma(j)}^{\alpha_{\sigma(j)}}
   \prod_{j=N-k+1}^N P_{\sigma(j)}^{3},\nonumber\\
 &=&\prod_{i=N-k+1}^N c_{0\,\sigma(i)}
   \prod_{i=1}^{N-k}a_{0\,\sigma(i)}
   \prod_{j=1}^{N-k}P_{\sigma(j)}^{\alpha_{\sigma(j)}}
   \prod_{j=N-k+1}^N P_{\sigma(j)}^{3},  \label{eq:T-tilde-2}
\end{eqnarray}
where $c_{0i}=c(u,z_i)$, $a_{0i}=a(u,z_i)$. Substituting
(\ref{eq:T-tilde-2}) into (\ref{eq:T-tilde-1}), we have
\begin{eqnarray}
F_{1\ldots N}T^{33}=\otimes_{i=1}^N
  \mbox{diag}\left(a_{0i},a_{0i},c_{0i}\right)_{(i)}
 F_{1\ldots N}.
\end{eqnarray}
Therefore,
\begin{eqnarray}
\tilde D(u)=\tilde T^{33}(u)=\otimes_{i=1}^N
  \mbox{diag}\left(a_{0i},a_{0i},c_{0i}\right)_{(i)}.\label{eq:T33-tilde}
\end{eqnarray}

The other elements of the monodromy matrix can then be obtained as
follows:
\begin{equation}
\tilde T^{3\alpha}=[\tilde E^{\alpha\,3},\tilde T^{33}],\quad
\tilde T^{\alpha 3}=[\tilde E^{3\,\alpha},\tilde T^{33}],\quad
(\alpha=1,2),
\end{equation}
which follows from the $gl(2|1)$ invariance of the $R$-matrix,
i.e. in terms of the monodromy matrix,
\begin{equation}
[\tilde T(u), \tilde E^{\alpha \beta}_{(0)}+\tilde E^{\alpha
\beta}]=0.
\end{equation}
 Substituting $\tilde
E^{\alpha\,3}$, $\tilde E^{3\,\alpha}$ and $\tilde T^{33}$ into
the above relations yields
\begin{eqnarray}
 \tilde T^{32}&=&-\sum_{i=1}^N b_{0i}E^{23}_{(i)}\otimes_{j\ne i}
   \mbox{diag}\left(a_{0j},a_{0j}a_{ij}^{-1},
             c_{0j}(2a_{ij})^{-1}\right)_{(j)},\label{eq:T32-F}\\
 \tilde T^{23}&=&\sum_{i=1}^N b_{0i}E^{32}_{(i)}\otimes_{j\ne i}
   \mbox{diag}\left(a_{0j},a_{0j},2c_{0j}\right)_{(j)},\\
 \tilde T^{31}&=&-\sum_{i=1}^N b_{0i}E^{13}_{(i)}\otimes_{j\ne i}
   \mbox{diag}\left(a_{0j}a_{ij}^{-1},a_{0j}a_{ij}^{-1},
             c_{0j}(2a_{ij})^{-1}\right)_{(j)}
              \nonumber\\ &&\mbox{}
   -\sum_{i\ne j=1}^N {a_{0i}b_{0j}\eta\over z_j-z_i}
     E^{12}_{(i)}\otimes E^{23}_{(j)}% \nonumber\\ &&
     \otimes_{k\ne i,j}
     \mbox{diag}\left(a_{0k}a_{ik}^{-1},a_{0k}a_{jk}^{-1},
               c_{0k}(2a_{jk})^{-1}\right)_{(k)},\label{eq:T31-F}
                \nonumber\\ &&\\
 \tilde T^{13}&=&\sum_{i=1}^N b_{0i}E^{31}_{(i)}\otimes_{j\ne i}
   \mbox{diag}\left(a_{0j},a_{0j}a_{ji}^{-1},2c_{0j}\right)_{(j)}
              \nonumber\\ &&\mbox{}
   +\sum_{i\ne j=1}^N {a_{0i}b_{0j}\eta\over z_i-z_j}
     E^{32}_{(i)}\otimes E^{21}_{(j)} %\nonumber\\ &&
     \otimes_{k\ne i,j}
     \mbox{diag}\left(a_{0k},a_{0k}a_{kj}^{-1},
               2c_{0k}\right)_{(k)}.
\end{eqnarray}
Here, $b_{0j}$ stands for $b(u,z_j)$.

\sect{Bethe vectors in the $F$-basis}
 ~~~Having obtained the creation operators of the $gl(2|1)$ monodromy
matrix in the $F$-basis, we are now in the position to study the
Bethe vectors in this basis. Acting the $F$-matrix on the
pseudo-vacuum state (\ref{de:vacuum}), we obtain
\begin{eqnarray}
F_{1\ldots N}|0>=\prod_{i<j}^N(1+c_{ij})|0>
 \equiv s(c)|0>.
\end{eqnarray}
Therefore, the $gl(2|1)$ Bethe vector (\ref{de:bs}) in the
$F$-basis can be written as
\begin{eqnarray}
   \tilde\Phi_N(v_1,\ldots,v_n)
 &\equiv& F_{1\ldots N}\Phi_N(v_1,\ldots,v_n)\nonumber\\
 &=& s(c)\sum_{d_1\ldots d_n}
  (\phi^{(1)}_n)^{d_1\ldots d_n}\tilde C_{d_1}(v_1)\ldots \tilde
  C_{d_n}(v_n)|0>, \label{eq:phi-F}
\end{eqnarray}
where $d_i=1,2$ $\,$ and\, $(\phi^{(1)}_n)^{d_1\ldots d_n}$ is the
coefficients of the nested Bethe vector\\
$\phi^{(1)}_n(v_1^{(1)},\ldots, v_m^{(1)})$, (\ref{de:nbs}),
associated with the $gl(2)$ transfer matrix $t^{(1)}(u)$ with
inhomogeneous parameters $v_1,\ldots,v_n$. The $c$-number
coefficient $(\phi^{(1)}_n)^{d_1\ldots d_n}$ has to be evaluated
in the original basis, not in the $F$-basis.

Let us first compute the $F$-transformed nested Bethe vector.
Denote by $F^{(1)}$ the $gl(2)$ $F$-matrix. Applying $F^{(1)}$ to
the nested Bethe vector $\phi^{(1)}_n$, we obtain
\begin{eqnarray}
  \tilde\phi^{(1)}_n(v_1^{(1)},\ldots, v_m^{(1)})
 &\equiv& F^{(1)}_{1\ldots n}
          \phi^{(1)}(v_1^{(1)},\ldots, v_m^{(1)}) \nonumber\\
 &=&\tilde C^{(1)}(v^{(1)}_1)
\tilde C^{(1)}(v^{(1)}_2)\ldots
           \tilde C^{(1)}(v^{(1)}_m)|0>^{(1)}, \label{eq:nbs-F}
\end{eqnarray}
where the nested pseudo-vacuum state $|0>^{(1)}$ is invariant
under the action of the $gl(2)$ $F$-matrix. From \cite{Maillet96},
the $F$-transformed $gl(2)$ creation operator $\tilde C^{(1)}$ is
given by
\begin{eqnarray}
 {\tilde C}^{(1)}(v^{(1)})=\sum_{i=1}^n
b(v^{(1)},v_i)\sigma_{(i)}^+ \otimes_{j\ne i}
\left(\begin{array}{cc} a(v^{(1)},v_j)a_{ij}^{-1}&0\\
0\quad\quad&1\end{array} \right)_{(j)}.
\end{eqnarray}
Substituting  ${\tilde C}^{(1)}(v)$ into (\ref{eq:nbs-F}), we
obtain
\begin{eqnarray}
 &&\tilde \phi^{(1)}_n(v_1^{(1)},\ldots,v_m^{(1)})
   =\tilde{C}^{(1)}(v_1^{(1)})\ldots \tilde{C}^{(1)}(v_m^{(1)})
   \,|0>^{(1)}\nonumber\\
&=&\sum_{i_1<\ldots< i_{m}}
B_{m}^{(1)}(v_1^{(1)},\ldots,v_{m}^{(1)}|v_{i_1},\ldots,v_{i_{m}})
\sigma_{(i_1)}^+\ldots\sigma_{(i_{m})}^+\,|0>^{(1)}\;,
\label{Psi_2}
\end{eqnarray}
where
\begin{eqnarray}
 B^{(1)}_m(v_1^{(1)},\ldots,v_m^{(1)}|v_{1},\ldots,v_{m})= \sum_{\sigma\in
S_m}\prod_{k=1}^m b(v_k^{(1)},v_{\sigma(k)})
\prod_{l=k+1}^{m}{{a(v_{k}^{(1)},v_{\sigma(l)})}
\over{a(v_{\sigma(k)},v_{\sigma(l)})}}.\nonumber\\
\label{eq:B_1}
\end{eqnarray}

Now back to the $gl(2|1)$ Bethe vector (\ref{eq:phi-F}). As is
shown in Appendix B, the Bethe vector is invariant (module overall
factor) under the exchange of arbitrary spectral parameters:
\begin{eqnarray}
\tilde\Phi_N(v_{\sigma(1)},\ldots,v_{\sigma(n)})
 =\mbox{sign}(\sigma)c^\sigma_{1\ldots n}
  \tilde\Phi_N(v_1,\ldots,v_n). \label{eq:exchange}
\end{eqnarray}
 This enable one to concentrate on a particularly simple term in the sum
(\ref{eq:phi-F}) of the following form with $p_1$ number of
$d_i=1$ and $n-p_1$ number of $d_j=2$
\begin{eqnarray}
 \tilde C_1(v_1)\ldots\tilde C_1(v_{p_1})
 \tilde C_2(v_{p_1+1})\ldots\tilde C_2(v_{n}).
% \equiv g_{1\ldots 12\ldots 2}(v_1,\ldots,p_1,p_{1}+1,\ldots n).
     \label{eq:C1-C2}
\end{eqnarray}
In the $F$-basis, the commutation relation between $C_i(v)$ and
$C_j(u)$, in (\ref{eq:commu}), becomes
\begin{eqnarray}
 \tilde C_i(v)\tilde C_j(u)
 &=&-{c(u,v)\over a(u,v)}\tilde C_j(u)\tilde C_i(v)
    -{b(u,v)\over a(u,v)}\tilde C_j(v)\tilde C_i(u).
    \label{eq:commu-CC-F}
\end{eqnarray}
Then using (\ref{eq:commu-CC-F}), all $\tilde C_1$'s in
(\ref{eq:C1-C2}) can be moved to the right of all $\tilde C_2$'s,
yielding
\begin{eqnarray}
 &&\tilde C_1(v_1)\ldots\tilde C_1(v_{p_1})
   \tilde C_2(v_{p_1+1})\ldots\tilde C_2(v_{n})= \nonumber\\
 &=&g(v_1,\ldots,v_n)
 \tilde C_2(v_{p_1+1})\ldots\tilde C_2(v_{n})
 \tilde C_1(v_{1})\ldots\tilde C_1(v_{p_1})+\ldots\ ,
 \label{eq:C2-C1}
\end{eqnarray}
where $g(v_1,\ldots,v_n)=\prod_{k=1}^{p_1}\prod_{l=p_1+1}^{n}
(-{c(v_l,v_k)/ a(v_l,v_k)})$ is the contribution from the first
term of (\ref{eq:commu-CC-F}) and ``$\ldots$" stands for the other
terms contributed by the second term of (\ref{eq:commu-CC-F}). It
is easy to see that the other terms have the form
\begin{equation}
\tilde C_2(v_{\sigma(p_1+1)})\ldots\tilde C_2(v_{\sigma(n)})\tilde
C_1(v_{\sigma(1)})\ldots\tilde C_1(v_{\sigma(p_1)})
\label{eq:C2-C1-sigma}
\end{equation}
with $\sigma\in {\cal S}_n$. Substituting (\ref{eq:C2-C1}) into
the Bethe vector (\ref{eq:phi-F}), we obtain
\begin{eqnarray}
\tilde\Phi_N^{p_1}(v_1,\ldots,v_n)%= \nonumber\\
 &=&s(c)(\phi^{(1)}_n)^{11\ldots 12\ldots 2}
    \prod_{k=1}^{p_1}\prod_{l=p_1+1}^{n}
    \left(-{c(v_{l},v_{k})
      \over a(v_{l},v_{k})}\right) \nonumber\\
&&\times\;{\tilde C}_{2}(v_{p_1 +1})\ldots {\tilde C}_{2}(v_{n})
{\tilde C}_{1}(v_{1})\ldots {\tilde C}_{1}(v_{p_1})|0> +\ldots\, ,
 \label{eq:Phi-1122}
\end{eqnarray}
where and below, we have used the up-index $p_1$ to denote the
Bethe vector corresponding to the quantum number $p_1$. All other
terms in (\ref{eq:Phi-1122}) (denoted as ``\ldots") are to be
obtained from the first term by the permutation (exchange)
symmetry. Thus (see Appendix C for the $n=2$ case),
\begin{eqnarray}
 &&{\tilde\Phi}_N^{p_1}(v_1,\ldots,v_n)= \nonumber\\
 &=&{s(c)\over p_1!(n-p_1)!}\sum_{\sigma \in {\cal S}_{n}}
     \mbox{sign}(\sigma)(c^{\sigma}_{1\ldots n})^{-1}
    (\phi^{(1),\sigma}_n)^{11\ldots 12\ldots 2}
    \prod_{k=1}^{p_1}\prod_{l=p_1+1}^{n}
    \left(-{c(v_{\sigma(l)},v_{\sigma(k)})
      \over a(v_{\sigma(l)},v_{\sigma(k)})}\right) \nonumber\\
&&\times\;{\tilde C}_{2}(v_{\sigma(p_1 +1)})\ldots {\tilde
C}_{2}(v_{\sigma(n)}) {\tilde C}_{1}(v_{\sigma(1)})\ldots {\tilde
C}_{1}(v_{\sigma(p_1)})\,|0>\, , \label{eq:Phi-phi}
\end{eqnarray}
where $(\phi^{(1),\sigma}_n)^{11\ldots 12\ldots 2}\equiv (\hat
f_\sigma\phi^{(1)}_n)^{11\ldots 12\ldots 2}$ with $\hat f_\sigma$
defined by (\ref{de:f-hat}) in the Appendix B.

We now show that $(\phi^{(1)}_n)^{1\ldots 12\ldots 2}$ in
(\ref{eq:Phi-phi}), which has to be evaluated in the original
basis, is invariant under the action of the $gl(2)$ $F$-matrix,
i.e.
\begin{equation}
 (\phi^{(1)}_n)^{11\ldots 12\ldots 2}
 =(\tilde\phi^{(1)}_n)^{11\ldots 12\ldots 2},
\end{equation}
so that it can be expressed in the form of (\ref{eq:B_1}).

Write the nested pseudo-vacuum vector in (\ref{de:nbs}) as
\begin{equation}
|0>^{(1)}\equiv |2\cdots 2>^{(1)},
\end{equation}
where the number of 2 is $n$. Then the nested Bethe vector
(\ref{de:nbs}) can be rewritten as
\begin{equation}
\phi^{(1)}_n(v_1^{(1)}\ldots v_{p_1}^{(1)})
 \equiv|\phi^{(1)}_n>
 =\sum_{d_1\ldots d_n}(\phi^{(1)}_n)^{d_1\ldots d_n}|d_1\ldots d_n>^{(1)}.
 \label{eq:phi1-phi1}
\end{equation}
Acting the $gl(2)$ $F$-matrix $F^{(1)}$ from left on the above
equation, we have
\begin{equation}
\tilde\phi^{(1)}_n(v_1^{(1)}\ldots v_{p_1}^{(1)})
 \equiv|\tilde\phi^{(1)}_n>=F^{(1)}|\phi^{(1)}_n>
 =\sum_{d_1\ldots d_n}(\tilde\phi^{(1)}_n)^{d_1\ldots d_n}|d_1\ldots d_n>^{(1)}
 .
 \label{eq:phi1-phi1-F}
\end{equation}
It follows that
\begin{eqnarray}
  (\tilde\phi^{(1)}_n)^{1\ldots 12\ldots2}
 &=&<1\ldots 12\ldots2|\tilde\phi^{(1)}_n>
  =<1\ldots 12\ldots2|F^{(1)}|\phi^{(1)}_n>\nonumber\\
 &=&<1\ldots 12\ldots2|\sum_{\sigma\in{\cal S}_n}
    \sum^{\quad\quad*}_{\alpha_{\sigma(1)}\ldots\alpha_{\sigma(n)}}\prod_{j=1}^n
    P_{\sigma(j)}^{\alpha_{\sigma(j)}}R^\sigma_{1\ldots n}
    |\phi^{(1)}_n> \label{eq:tildephi-phi-1}\\
 &=&<1\ldots 12\ldots2|\left.\left\{
    \sum^{\quad\quad*}_{\alpha_{\sigma(1)}\ldots\alpha_{\sigma(n)}}\prod_{j=1}^n
    P_{\sigma(j)}^{\alpha_{\sigma(j)}}\right\}
    \right|_{\sigma=id}R^{\sigma=id}_{1\ldots n}
    |\phi^{(1)}_n> \label{eq:tildephi-phi-2}     \\
 &=&<1\ldots 12\ldots2|\phi^{(1)}_n>
  =(\phi^{(1)}_n)^{1\ldots 12\ldots2}, \label{eq:tildephi-phi-3}
\end{eqnarray}

Summarizing, we propose the following form of the $gl(2|1)$ Bethe
vector
\begin{eqnarray}
 &&{\tilde\Phi}_N^{p_1}(v_1,\ldots,v_n)= \nonumber\\
 &=&{s(c)\over p_1!(n-p_1)!}\sum_{\sigma \in {\cal S}_{n}}
    \mbox{sign}(\sigma) (c^{\sigma}_{1\ldots n})^{-1}
    B^{(1)}_{p_1}(v_{1}^{(1)},\ldots,v_{p_1}^{(1)}|
    v_{\sigma(1)},\ldots,v_{\sigma(p_1)}) \nonumber\\
&&\times\;
    \prod_{k=1}^{p_1}\prod_{l=p_1+1}^{n}
    \left(-{c(v_{\sigma(l)},v_{\sigma(k)})
      \over a(v_{\sigma(l)},v_{\sigma(k)})}\right)
    {\tilde C}_{2}(v_{\sigma(p_1 +1)})\ldots {\tilde C}_{2}(v_{\sigma(n)})
\nonumber\\ &&\times\;
    {\tilde C}_{1}(v_{\sigma(1)})\ldots {\tilde
C}_{1}(v_{\sigma(p_1)})\,|0>\, .  \label{eq:Phi-3a}
\end{eqnarray}
Substituting (\ref{eq:T32-F}) and (\ref{eq:T31-F}) into the above
relation, we finally obtain
\begin{eqnarray}
&&{\tilde\Phi}_N^{p_1}(v_1,\ldots,v_n)\nonumber\\
 &=&{s(c)\over p_1!(n-p_1)!}
 \sum_{{i_1<\ldots<i_{p_1}}}\sum_{{i_{p_1+1}<\ldots<i_{n}}}
  B_{n,p_1}(v_1,\ldots,v_{n};v_{1}^{(1)},\ldots,
              v_{p_1}^{(1)}|z_{i_1},\ldots,z_{i_{n}})\nonumber\\
&&\times  \prod_{j={p_1+1}}^{{n}}
E^{23}_{(i_j)}\prod_{j={1}}^{{p_1}}E^{13}_{(i_j)}\,|0>,
\end{eqnarray}
where
$\{i_1,i_2,\ldots,i_{p_1}\}\cap\{i_{p_1+1},i_{p_1+2},\ldots,i_{n}\}
 =\varnothing
$ and
\begin{eqnarray}
 &&B_{n,p_1}(v_1,\ldots,v_{n};v_{1}^{(1)},\ldots,
              v_{p_1}^{(1)}|z_{i_{1}},\ldots,z_{i_{n}})=\nonumber\\
 &=&\sum_{\sigma \in S_{n}}
    \mbox{sign}(\sigma)(c^{\sigma}_{1\ldots n})^{-1}
  \prod_{k=1}^{p_1}\prod_{l=p_1+1}^{n}
    \left(-{c(v_{\sigma(l)},v_{\sigma(k)})a(v_{\sigma(l)},z_{i_k})
                          \over a(v_{\sigma(l)},v_{\sigma(k)})}\right)
      \nonumber\\ && \times
         B_{n-p_1}^*(v_{\sigma(p_1+1)},\ldots,
                      v_{\sigma(n)}|z_{i_{p_1+1}},\ldots,z_{i_{n}})
                       \nonumber\\
&&\times
B_{p_1}^{(1)}(v_{1}^{(1)},\ldots,v_{p_1}^{(1)}|v_{\sigma(1)},\ldots,
                       v_{\sigma(p_1)})
         B_{p_1}^*(v_{\sigma(1)},\ldots,
                       v_{\sigma(p_1)}|z_{i_{1}},\ldots,z_{i_{p_1}})\nonumber\\
\label{B1}
\end{eqnarray}
with
\begin{eqnarray}
 && B^*_p(v_1,\ldots,v_p|z_{1},\ldots,z_{p})=\nonumber\\
 &=& \sum_{\sigma\in {\cal S}_p} \mbox{sign}(\sigma)
 \prod_{m=1}^p (-b(v_m,z_{\sigma(m)}))
    \prod_{j\ne\sigma(p),\ldots,\sigma(m)}^N{c(v_m,z_j)\over 2a(z_{\sigma(m)},z_j)}
\prod_{l=m+1}^{p}{{a(v_{m},z_{\sigma(l)})}\over{a(z_{\sigma(m)},z_{\sigma(l)})}}.\nonumber\\
\label{eq:B*}
\end{eqnarray}

\section{Discussions}
~~~ In this paper, we have constructed the factorizing
$F$-matrices for the supersymmetric $t$-$J$ model. In the basis
provided by the $F$-matrix (the $F$-basis), the monodromy matrix
and the creation operators take completely symmetric forms. We
moreover have obtained a simple representation of the Bethe vector
of the system.

Authors in \cite{Korepin99} solved the quantum inverse problem of
the supersymmetric $t$-$J$ model in the original basis. Namely
they reconstructed the local operators ($E^{ij}$) in terms of
operators figuring in the $gl(2|1)$ monodromy matrix. This
together with the results of the present paper in the $F$-basis
should enable one to get the exact representations of form factors
and correlation functions of the supersymmetric $t$-$J$ model.
These are under investigation and
results will be reported elsewhere.\\[5mm]
{\bf Acknowledgements:} The authors would like to thank Vladimir
Korepin for discussions and communications in the early stage of
the work.

This work was financially supported by the Australia Research
Council. S.Y. Zhao has also been supported by the UQ Postdoctoral
Research Fellowship.

\section*{Appendix A $\quad$ The nested Bethe ansatz for the $gl(2|1)$ model}
\setcounter{equation}{0}
\renewcommand{\theequation}{A.\arabic{equation}}
~~~ In this Appendix, we recall the nested Bethe ansatz method
\cite{Ess92}\cite{Foer931}. The Hamiltonian (\ref{de:H}) can be
exactly diagonalized by using the nested Bethe ansatz method.
Define the pseudo-vacuum state
\begin{eqnarray}
|0>_k=\left(\begin{array}{c} 0\\0\\1\end{array}\right),~~~~
|0>=\otimes _{k=1}^N|0>_k\label{de:vacuum}
\end{eqnarray}
and  the Bethe vector
\begin{eqnarray}
 \Phi_N(v_1,\ldots,v_n)=\sum_{d_1,\ldots,d_n}
 (\phi^{(1)}_n)^{d_1\ldots d_n} C_{d_1}(v _1) C_{d_2}(v _2)
  \ldots C_{d_n}(v _n)|0>,
\label{de:bs}
\end{eqnarray}
where $(\phi^{(1)}_n)^{d_1\ldots d_n}$ is a function of the
spectral parameters $v _j$.

Applying the quantum operators $A_{ij},B_i,C_i$ and $D$ to the
pseudo-vacuum state, we obtain
\begin{eqnarray}
&&D(u)|0>=\prod_{i=1}^N c(u,z_i)|0>,\quad \quad
A_{ij}(u)|0>=\delta_{ij}\prod_{k=1}^N a(u,z_k)|0>, \nonumber\\
&& B_i|0>=0.
\end{eqnarray}
From the GYBE (\ref{eq:GYBE}), one obtains the following
commutation relations
\begin{eqnarray}
C_i(u)C_j(v)&=&-\sum_{k,l}{r(u,v)^{kl}_{ji}\over c(u,v)}C_k(v)C_l(u),\nonumber\\
%C_i(v)C_j(u)&=&-{c(u,v)\over a(u,v)}C_j(u)C_i(v)
%             -{b(u,v)\over a(u,v)}C_j(u)C_i(v), \nonumber\\
D(u)C_j(v)&=&{c(v,u)\over a(v,u)}C_j(v)D(u)
             +{b(v,u)\over a(v,u)}C_j(u)D(v), \nonumber\\
A_{ij}(u)C_k(v)&=&\sum_{m,l}{r(u,v)^{ml}_{jk}\over
a(u,v)}C_l(v)A_{im}(u)
             -{ b(u,v)\over a(u,v)}C_j(u)A_{ik}(v),
             \label{eq:commu}
\end{eqnarray}
where
\begin{eqnarray}
r(u,v)=\left(\begin{array}{cccc}
 1&0&0&0\\ 0&a(u,v)&b(u,v)&0\\ 0&b(u,v)&a(u,v)&0\\ 0&0&0&1
           \end{array}\right) \label{de:r}
\end{eqnarray}
is the $gl(2)$ $R$-matrix acting on the tensor product of the
2-dimensional representation of $gl(2)$. With the help of the
commutation relations (\ref{eq:commu}), we have the action of
$D(u)$ on the Bethe vector
\begin{eqnarray}
&&D(u)\sum_{d_1,\ldots,d_n}(\phi^{(1)}_n)^{d_1\ldots d_n}
C_{d_1}(v _1) C_{d_2}(v_2)
     \ldots C_{d_n}(v _n)|0> \nonumber\\
&=&\prod_{i=1}^N c(u,z_i)\prod_{j=1}^n
    {c(v_j,u)\over a(v_j,u)}\sum_{d_1,\ldots,d_n}(\phi^{(1)}_n)^{d_1\ldots d_n}
    C_{d_1}(v _1)\ldots C_{d_n}(v _n)|0> +u.t. ,
    \label{eq:D}
\end{eqnarray}
Similarly, the action of $A_{aa}$ ($a=1,2$) on the Bethe vector
gives rise to
\begin{eqnarray}
&&A_{aa}(u)\sum_{d_1,\ldots,d_n}(\phi^{(1)}_n)^{d_1\ldots d_n} C_{d_1}(v _1)
     \ldots C_{d_n}(v _n)|0> \nonumber\\
&=&\prod_{i=1}^Na(u,z_i)\prod_{j=1}^n{1\over a(u,v_j)}
   \sum_{d_1,\ldots,d_n}(\phi^{(1)}_n)^{d_1\ldots d_n} C_{q_1}(v _1)
   \ldots C_{q_n}(v _n)|0>\nonumber\\ && \times
     r(u,v_1)^{c_1q_1}_{ad_1}\ r(u,v_1)^{c_2q_2}_{c_1d_2}\ldots
     r(u,v_n)^{\ \ a\ \ q_n}_{c_{n-1}d_n}+u.t. \nonumber\\
&\equiv&\prod_{i=1}^Na(u,z_i)\prod_{j=1}^n{1\over a(u,v_j)}
    C_{q_1}(v_1)\ldots C_{q_n}(v _n)|0>\nonumber\\ && \times
    \sum_{d_1,\ldots,d_n}t^{(1)}(u)^{q_1\ldots q_n}_{d_1\ldots d_n}
    (\phi^{(1)}_n)^{d_1\ldots d_n} +u.t., \label{eq:A}
\end{eqnarray}
where
\begin{equation}
t^{(1)}(u)=tr_0T^{(1)}(u)     \label{de:tr1}
\end{equation}
is the nested transfer matrix with
\begin{eqnarray}
T^{(1)}(u)&\equiv& r_{n}(u,v_n)\ldots r_{1}(u,v_1) \nonumber\\
&=&L_{n}^{(1)}(u,v_{n})\ldots
  L_{i}^{(1)}(u,v_{i})\ldots L_1^{(1)}(u,v_1) \nonumber\\
&=&\left(\begin{array}{cc} A^{(1)}(u)&B^{(1)}(u)\\
C^{(1)}(u)&D^{(1)}(u)\end{array}\right)
 \label{de:T1}
\end{eqnarray}
being the nested monodromy matrix. (\ref{eq:A}) results in an
eigenvector of $A_{aa}(u)$ if
 $\sum_{d_1,\ldots,d_n}t^{(1)}(u)^{q_1\ldots q_n}_{d_1\ldots d_n}
  (\phi^{(1)}_n)^{d_1\ldots d_n}
  =\varepsilon^{(1)}(u)(\phi^{(1)}_n)^{q_1\ldots q_n}.$ This is
  nothing but a Bethe ansatz problem for $gl(2)$ chain of length
  $n$ with the inhomogeneities now given by the parameters
  $v_1,\ldots, v_n$ of the $gl(2|1)$ problem. This inspires one to
  define $\phi^{(1)}_n$ as
\begin{equation}
\phi^{(1)}_n(v_1^{(1)},\ldots, v_m^{(1)})=C^{(1)}(v^{(1)}_1)
C^{(1)}(v^{(1)}_2)\ldots
           C^{(1)}(v^{(1)}_m)|0>^{(1)}, \label{de:nbs}
\end{equation}
where $|0>^{(1)}$ is the 2-dimensional pseudo-vacuum
\begin{eqnarray}
|0>^{(1)}_k=\left(\begin{array}{c} 0\\1\end{array}\right),~~~~
|0>^{(1)}=\otimes _{k=1}^n|0>^{(1)}_k.    \label{de:n-vacuum}
\end{eqnarray}
Then $\phi^{(1)}_n$ spans a subspace of the space spanned by
$\Phi$.

The action of the nested monodromy matrix elements on the nested
pseudo-vacuum (\ref{de:n-vacuum}) read
\begin{eqnarray}
&&A^{(1)}(u)|0>^{(1)}=\prod_{i=1}^n a(u,v_i)|0>^{(1)},\quad
  D^{(1)}(u)|0>^{(1)}=|0>^{(1)},\quad \nonumber\\
&&B^{(1)}(u)|0>^{(1)}=0.
\end{eqnarray}
The commutation relations between the elements of the nested
monodromy matrix are given
\begin{eqnarray}
A^{(1)}(u)C^{(1)}(v)&=&{1\over a(u,v)}C^{(1)}(v)A^{(1)}(u)
     -{b(u,v)\over a(u,v)}C^{(1)}(u)(u)A^{(1)}(v), \nonumber\\
D^{(1)}(u)C^{(1)}(v)&=&{1\over a(v,u)}C^{(1)}(v)D^{(1)}(u)
     -{b(v,u)\over a(v,u)}C^{(1)}(u)D^{(1)}(v), \nonumber\\
C^{(1)}(u)C^{(1)}(v)&=&C^{(1)}(v)C^{(1)}(u).
\end{eqnarray}
Applying the nested transfer matrix (\ref{de:tr1}) to the nested
Bethe state (\ref{de:nbs}), one obtains the eigenvalue of the
nested system:
\begin{eqnarray}
\varepsilon^{(1)}(u)=\prod_{i=1}^n a(u,v_i)
   \prod_{j=1}^m{1\over a(u,v_j^{(1)})}
    +\prod_{j=1}^m{1\over a(v_j^{(1)},u)},
\end{eqnarray}
where $v_j^{(1)}$ is constrained by the nested Bethe ansatz
equations:
\begin{eqnarray}
\prod_{\alpha=1,\ne \beta}^m
   {v_\beta^{(1)}-v_\alpha^{(1)}+\eta\over
    v_\beta^{(1)}-v_\alpha^{(1)}-\eta}
=\prod_{\gamma=1}^n {v_\beta^{(1)}-v_\gamma-{\eta\over 2}\over
v_\beta^{(1)}-v_\gamma+{\eta\over 2}} \quad\quad
(\beta=1,2,\dots,m).
\end{eqnarray}

Then from (\ref{de:t}) and (\ref{eq:D}-\ref{eq:A}), we obtain the
eigenvalue $\varepsilon(u)$ of the supersymmetric $t$-$J$ model:
$t(u)\Phi=\varepsilon(u)\Phi$ with
\begin{eqnarray}
\varepsilon(u)&=&
 \prod_{i=1}^Na(u,z_i)\prod_{j=1}^n{1\over a(u,v_j)}
  \left(\prod_{i=1}^n a(u,v_i)
   \prod_{j=1}^m{1\over a(u,v_j^{(1)})}
    +\prod_{j=1}^m{1\over a(v_j^{(1)},u)}\right)\nonumber\\
&&\mbox{}    -\prod_{i=1}^N c(u,z_i)\prod_{j=1}^n
    {c(v_j,u)\over a(v_j,u)}.
\end{eqnarray}
Here the unwanted terms in (\ref{eq:D}-\ref{eq:A}) vanish when we
take supertrace of the transfer matrix, which yields the Bethe
ansatz equations
\begin{eqnarray}
\prod_{i=1}^N{v_\beta-z_i-\eta\over
                          v_\beta-z_i}
\prod_{\alpha=1,\ne\beta}^n{v_\alpha-v_\beta-\eta\over
                            v_\alpha-v_\beta+\eta}
\prod_{\gamma=1}^m{v^{(1)}_\gamma-v_\beta\over
                            v^{(1)}_\gamma-v_\beta+\eta}
=1 \quad\quad (\beta=1,2,\dots,m).
\end{eqnarray}

\section*{Appendix B $\quad$ The exchange symmetry of the Bethe vector}
\setcounter{equation}{0}
\renewcommand{\theequation}{B.\arabic{equation}}

~~~~For the Bethe vector $\Phi_N(v_1,\ldots,v_n)$ of the
supersymmetric $t$-$J$ model, we define the exchange operator
$\hat f_\sigma=\hat f_{\sigma_1}\ldots \hat f_{\sigma_k}$ by
\begin{eqnarray}
 \hat f_\sigma \Phi_N(v_1,v_2,\ldots,v_n)
 =\Phi_N(v_{\sigma(1)},v_{\sigma(2)},\ldots,v_{\sigma(n)}),
  \label{de:f-hat}
\end{eqnarray}
where $\sigma\in {\cal S}_n$ and $\sigma_i$ are elementary
permutations. In \cite{Vega89}\cite{Tak83}\cite{Albert01}, it has
been shown that the $gl(m)$ Bethe vector is invariant under the
action of the exchange operator $\hat f_\sigma$. In this appendix,
we examine the exchange symmetry of the $gl(2|1)$ Bethe vector.

We first study the exchange symmetry for the elementary exchange
operator $\hat f_{\sigma_i}$ which exchanges the parameter $v_i$
and $v_{i+1}$. Acting $\hat f_{\sigma_i}$ on the Bethe vector of
$gl(2|1)$ (\ref{eq:phi-F}), we have
\begin{eqnarray}
&&\hat f_{\sigma_i} \Phi_N(v_1,v_2,\ldots,v_n)%\nonumber\\
 =\Phi_N(v_1,\ldots,v_{i+1},v_i,\ldots,v_{n})\nonumber\\
 &=&\sum_{d_1,\ldots,d_n}
 (\phi^{(1),\sigma_i}_n)^{d_1\ldots d_n} C_{d_1}(v _1)\ldots C_{d_i}(v_{i+1})
 C_{d_{i+1}}(v_i)\ldots C_{d_n}(v _n)|0>, \label{eq:f-Phi}
\end{eqnarray}
where $(\phi^{(1),\sigma_i}_n)^{d_1\ldots d_n}$ is constructed by
the nested monodromy matrix
\begin{eqnarray}
T^{(1),\sigma_i}(u)&=&L_{n}^{(1)}(u,v_{n})\ldots
L_{i+1}^{(1)}(u,v_{i})
  L_{i}^{(1)}(u,v_{i+1})\ldots L_1^{(1)}(u,v_1).
\label{eq:mono-nest-i}
\end{eqnarray}

The commutation relation between $C_i$ and $C_j$ in
(\ref{eq:commu}) can be rewritten as
\begin{eqnarray}
C_i(u)C_j(v)&=&-\sum_{k,l}{\check r(u,v)^{kl}_{ij}\over
c(u,v)}C_k(v)C_l(u) \label{eq:commu-cc-b}
\end{eqnarray}
by using the braided $r$-matrix $\check{r}(u,v)\equiv {\cal
P}r(u,v)$, where ${\cal P}$ permutes the tensor spaces of the
2-dimensional $gl(2)$-module. Then, by (\ref{eq:commu-cc-b}),
(\ref{eq:f-Phi}) becomes
\begin{eqnarray}
\hat f_{\sigma_i} \Phi_N(v_1,v_2,\ldots,v_n)%\nonumber\\
&=&-c(v_i,v_{i+1})\sum_{d_1,\ldots,d_n}
 (\phi^{(1),\sigma_i}_n)^{d_1\ldots d_n} C_{d_1}(v_1)\ldots
 \nonumber\\ &&\times
 (\check{r}(v_{i+1},v_i))^{k\ \ l}_{d_id_{i+1}}
 C_{k}(v_i)C_{l}(v_{i+1})\ldots
 C_{d_{n}}(v_{n})|0>.\label{eq:f-Phi-r}
\end{eqnarray}
We now compute the action of  $(\check r(v_{i+1},v_i))^{k\ \
l}_{d_id_{i+1}}$ on $(\phi^{(1),\sigma_i})^{d_1\ldots d_n}$. One
checks that $\check r$-matrix satisfies the YBE
\begin{eqnarray}
&&\check r_{i\,i+1}(v_{i+1},v_{i})
  L_{i+1}^{(1)}(u,v_{i})L_{i}^{(1)}(u,v_{i+1})\nonumber\\
&&\hspace{2em}=L_{i+1}^{(1)}(u,v_{i+1})L_{i}^{(1)}(u,v_{i})
  \check{r}_{i\,i+1}(v_{i+1},v_{i})\,.
\end{eqnarray}
Therefore, acting $\check r$ on $T^{(1),\sigma_i}(u)$, we have
\begin{eqnarray}
 \check{r}_{i\,i+1}(v_{i+1},v_{i})T^{(1),\sigma_i}(u)
 =T^{(1)}(u)\check{r}_{i\,i+1}(v_{i+1},v_{i}) .
\end{eqnarray}
Thus, because $\check r\, v_2\otimes v_2=v_2\otimes v_2$, we
obtain
\begin{eqnarray}
 (\phi^{(1)}_n)^{d_1\ldots kl\ldots d_n}
 =\sum_{d_i d_{i+1}}(\check r(v_{i+1},v_i))_{d_id_{i+1}}^{k\,\,\,l}
 (\phi^{(1),\sigma_i}_n)^{d_1\ldots d_i d_{i+1}\ldots  d_n}.
\end{eqnarray}
Changing the indices $k, l$ to $d_i,d_{i+1}$, respectively, and
substituting the above relation into (\ref{eq:f-Phi-r}), we obtain
the exchange symmetric relation of the Bethe vector of $gl(2|1)$
\begin{eqnarray}
\hat f_{\sigma_i} \Phi_N(v_1,v_2,\ldots,v_n)%\nonumber\\
&=&-c(v_i,v_{i+1})\Phi_N(v_1,v_2,\ldots,v_n)
\end{eqnarray}
for the elementary permutation operator $\sigma_i$.

It follows that under the action of the exchange operator
$f_{\sigma}$
\begin{eqnarray}
  \hat f_{\sigma} \Phi_N(v_1,v_2,\ldots,v_n)%\nonumber\\
 &=&\mbox{sign}(\sigma)c^\sigma_{1\ldots n}
  \Phi_N(v_1,v_2,\ldots,v_n),
\end{eqnarray}
where $\mbox{sign}(\sigma)=1$ if $\sigma$ is even and
$\mbox{sign}(\sigma)=-1$ if $\sigma$ is odd, and
$c^\sigma_{1,\ldots,n}$ has the decomposition law
\begin{eqnarray}
 c^{\sigma'\sigma}_{1\ldots n}
 =c^{\sigma}_{\sigma'(1\ldots n)}
  c^{\sigma'}_{1\ldots n}\label{eq:c-cc}
\end{eqnarray}
with $c^{\sigma_i}_{1\ldots n}=c_{i\ i+1}\equiv c(v_i,v_{i+1})$
for an elementary permutation $\sigma_i$.

\section*{Appendix C $\quad$ An example of (\ref{eq:Phi-phi}) }
\setcounter{equation}{0}
\renewcommand{\theequation}{C.\arabic{equation}}

~~~~ As an illustration, we give a detailed derivation of
(\ref{eq:Phi-phi}) for the $n=2$ case. In the $F$-basis, the
$gl(2|1)$ Bethe vector is given by
\begin{eqnarray}
\tilde\Phi_N(v_1,v_2)=s(c)\sum_{d_1,d_2}
  (\phi^{(1)}_2)^{d_1d_2}\tilde C_{d_1}(v_1)\tilde C_{d_2}(v_2)|0>
\end{eqnarray}
with $d_1,d_2=1,2$. The quantum number $p_1$ may take three values
0,1 or 2. Here we concentrate on the $p_1=1$ case and the
$p_1=0,2$ cases can be treated similarly. We have
\begin{eqnarray}
{1\over
s(c)}\tilde\Phi_N^{p_1=1}(v_1,v_2)&=&(\phi^{(1)}_2)^{12}\tilde
 C_{1}(v_1)\tilde C_{2}(v_2)|0>
 +(\phi^{(1)}_2)^{21}\tilde C_{2}(v_1)\tilde C_{1}(v_2)|0>\nonumber\\
 &=&g(v_1,v_2)(\phi^{(1)}_2)^{12}\tilde C_{2}(v_2)\tilde C_{1}(v_1)|0> \nonumber\\
 &&\mbox{}
 +\left[g'(v_1,v_2)(\phi^{(1)})^{12}
 +(\phi^{(1)}_2)^{21}\right]\tilde C_{2}(v_1)\tilde C_{1}(v_2)|0>,  \label{eq:phi-2}
   \end{eqnarray}
where $g(v_1,v_2)=-c(v_2,v_1)/a(v_2,v_1)$,
$g'(v_1,v_2)=-b(v_2,v_1)/a(v_2,v_1)$. Acting $\hat f_{\sigma_1}$
on (\ref{eq:phi-2}), we have
\begin{eqnarray}
 {1\over s(c)}\hat f_{\sigma_1}\tilde\Phi_N^1(v_1,v_2)
 &=&{1\over s(c)}\tilde\Phi_N^1(v_2,v_1)\nonumber\\
 &=&g(v_2,v_1)(\phi^{(1),\sigma_1}_2)^{12}\tilde C_{2}(v_1)\tilde C_{1}(v_2)|0>. \nonumber\\
 &&\mbox{}
 +\left[g'(v_2,v_1)(\phi^{(1),\sigma_1}_2)^{12}
 +(\phi^{(1),\sigma_1}_2)^{21}\right]\tilde C_{2}(v_2)\tilde C_{1}(v_1)|0>,  \label{eq:phi-2-f}
\end{eqnarray}
Now the exchange symmetry
$$\tilde\Phi_N^1(v_2,v_1)=-c(v_1,v_2)\tilde\Phi_N^1(v_1,v_2)
$$
gives rise to the relation:
\begin{eqnarray}
-c(v_2,v_1)g(v_2,v_1)(\phi^{(1),\sigma_1}_2)^{12}&=&g'(v_1,v_2)(\phi^{(1)}_2)^{12}
 +(\phi^{(1)}_2)^{21}. \label{eq:phi12-phi21}
\end{eqnarray}
By means of (\ref{eq:phi12-phi21}), one may recast
(\ref{eq:phi-2}) into the form
\begin{eqnarray}
{1\over s(c)}\tilde\Phi_N^1(v_1,v_2)
 &=&g(v_1,v_2)(\phi^{(1)}_2)^{12}\tilde C_{2}(v_2)\tilde C_{1}(v_1)|0> \nonumber\\
 &&\mbox{}
 -c(v_2,v_1)g(v_2,v_1)(\phi^{(1),\sigma_1}_2)^{12}\tilde C_{2}(v_1)\tilde C_{1}(v_2)|0>,
\end{eqnarray}
which coincides with (\ref{eq:Phi-phi}) for $n=2$.

 \end{document}